\documentclass[preprint, 3p, times, lefttitle]{elsarticle} 

\usepackage{comment}
\usepackage{silence} 
\usepackage{blindtext}
\usepackage{xurl}
\usepackage{multirow}
\usepackage{enumitem}
\usepackage{amssymb}
\usepackage{siunitx}
\usepackage{bm}
\usepackage{arydshln}
\usepackage{float}
 
\usepackage{pgfplots}
\usepackage{tikz}
\pgfplotsset{compat=1.17} 
\usepackage[mode=buildnew]{standalone}
\usetikzlibrary{decorations.pathreplacing,calligraphy}
\usepackage{pgfplots}
\usetikzlibrary{pgfplots.dateplot}
    \usepgfplotslibrary{dateplot}

\definecolor{color1}{RGB}{0,0,0} 
\definecolor{color2}{RGB}{182,193,251} 
\definecolor{bblue}{HTML}{4F81BD}
\definecolor{rred}{HTML}{C0504D}
\definecolor{ggreen}{HTML}{9BBB59}
\definecolor{ppurple}{HTML}{9F4C7C}
\definecolor{sandybrown}{HTML}{f4a460}
\definecolor{lightseagreen}{HTML}{20b2aa}
\definecolor{cornflowerblue}{HTML}{6495ed}
\definecolor{limegreen}{HTML}{32CD32}
\definecolor{orange}{HTML}{ffa500}
\definecolor{purple}{HTML}{6495ed}
\definecolor{mediumaquamarine}{HTML}{66cdaa}
\definecolor{dimgray}{HTML}{696969}
\definecolor{rblue}{HTML}{006795}
\definecolor{rorange}{HTML}{F26035}

\tikzstyle{Tank} = [rectangle, rounded corners, minimum width=1cm, minimum height=0.7cm,text centered, draw=black, fill=cornflowerblue!10] 
\tikzstyle{arrow} = [very thin,->,>=stealth]
\tikzstyle{line} = [very thin,-,>=stealth]
\tikzstyle{blue_arrow} = [very thin,<->,>=stealth, draw=bblue]

\usepackage[acronym, automake]{glossaries}
\makeglossaries
\newacronym{ICS}{ICS}{industrial control system}
\newacronym{PLC}{PLC}{programmable logic controller}
\newacronym{HMI}{HMI}{human-machine interface}
\newacronym{IDS}{IDS}{intrusion detection systems}
\newacronym{ML}{ML}{machine learning}
\newacronym{CPU}{CPU}{central processing unit}
\newacronym{SCADA}{SCADA}{supervisory control and data acquisition}
\newacronym{SAS}{SAS}{substation automation system}
\newacronym{DSO}{DSO}{distribution system operator}
\newacronym{DoS}{DoS}{denial-of-service}
\newacronym{MITM}{MITM}{man-in-the-middle}
\newacronym{SVM}{SVM}{support vector machine}
\newacronym{ANN}{ANN}{artificial neural network}
\newacronym{CNN}{CNN}{convolutional neural network}
\newacronym{RF}{RF}{random forest}
\newacronym{KNN}{KNN}{k-nearest neighbors}
\newacronym{GB}{GB}{gradient boosting}
\newacronym{SMOTE}{SMOTE}{synthetic minority oversampling technique}
\newacronym{PCA}{PCA}{principal component analysis}
\newacronym{IHT}{IHT}{instance hardness threshold}
\newacronym{CV}{CV}{cross-validation}
\newacronym{CPS}{CPS}{cyber-physical system}
\newacronym{KPI}{KPI}{key performance indicator}
\newacronym{IPFIX}{IPFIX}{Internet Protocol Flow Information Export}
\newacronym{FP}{FP}{false positive}
\newacronym{TP}{TP}{true positive}
\newacronym{LSTM}{LSTM}{long short-term memory}
\newacronym{FS}{FS}{flow sensor}
\newacronym{mPMU}{PMU}{phasor measurement unit}
\newacronym{SIEM}{SIEM}{Security Information and Event Management}

\def\BibTeX{{\rm B\kern-.05em{\sc i\kern-.025em b}\kern-.08em
    T\kern-.1667em\lower.7ex\hbox{E}\kern-.125emX}}
    
\usepackage{stfloats}
\biboptions{numbers,sort&compress}
\journal{Computers in Industry}

\begin{document}
\fontsize{10}{11}\selectfont
\begin{frontmatter}

\title{CyPhERS: A Cyber-Physical Event Reasoning System providing real-time situational awareness for attack and fault response}

\cortext[cor1]{Corresponding author. E-mail address: nilmu@dtu.dk (N. Müller).}

\author[inst1]{Nils Müller\corref{cor1}}
\author[inst2]{Kaibin Bao}
\author[inst2]{Jörg Matthes}
\author[inst1]{Kai Heussen}

\affiliation[inst1]{organization={Wind and Energy Systems Department, Technical University of Denmark},
            addressline={Building 330, Risø campus}, 
            city={4000 Roskilde},
            country={Denmark}}

\affiliation[inst2]{organization={Institute for Automation and Applied Informatics, Karlsruhe Institute of Technology},
            addressline={Building 445, Campus North}, 
            city={76344 Eggenstein-Leopoldshafen},
            country={Germany}}

\begin{abstract}
\Glspl{CPS} constitute the backbone of critical infrastructures such as power grids or water distribution networks. 
Operating failures in these systems can cause serious risks for society. 
To avoid or minimize downtime, operators require real-time awareness about critical incidents.
However, online event identification in \glspl{CPS} is challenged by the complex interdependency of numerous physical and digital components, requiring to take cyber attacks and physical failures equally into account.
The online event identification problem is further complicated through the lack of historical observations of critical but rare events, and the continuous evolution of cyber attack strategies. 
This work introduces and demonstrates CyPhERS, a \textbf{Cy}ber-\textbf{Ph}ysical \textbf{E}vent \textbf{R}easoning \textbf{S}ystem. 
CyPhERS provides real-time information pertaining the occurrence, location, physical impact, and root cause of potentially critical events in \glspl{CPS}, without the need for historical event observations.
Key novelty of CyPhERS is the capability to generate informative and interpretable event signatures of known and unknown types of both cyber attacks and physical failures. 
The concept is evaluated and benchmarked on a demonstration case that comprises a multitude of attack and fault events targeting various components of a \gls{CPS}.
The results demonstrate that the event signatures provide relevant and inferable information on both known and unknown event types. 
\end{abstract}


\end{frontmatter}
\glsresetall
\section{Introduction} \label{sec:introduction}
The recent development of critical infrastructure such as power grids and water distribution networks is driven by digitalization and automation. 
The closed-loop integration of physical processes with computer systems and communication technologies renders them \glspl{CPS}.
In such systems, critical incidents can arise from failures of a variety of interconnected physical and digital devices \cite{ALGULIYEV2018212, COLABIANCHI2021107534}.
The increasing trend of connecting critical infrastructure to the internet adds cyber attacks as another dimension of possible incident causes \cite{MAGLARAS201842}. 
Cyber attacks against \glspl{CPS} constitute a particular risk, as they can entail damage to physical equipment or even humans.
Appropriate countermeasures for critical incidents are facilitated by real-time information about affected devices, root causes and physical impact.
As incidents can be caused by failure or attack against a variety of physical and digital devices, integrated monitoring of the cyber and physical domain is advantageous \cite{8893316}. 
The problem is further complicated by new attack types or unseen physical failures, where no prior knowledge is available for event identification. 
Consequently, a monitoring system for \glspl{CPS} is required which provides real-time information about unknown and known types of both cyber attacks and physical failures. 

Nowadays, monitoring of the cyber and physical domain is largely conducted in isolated silos, for example through intrusion or fault detection systems. 
Some recent works propose integration of both through supervised \gls{ML} \cite{muller2022assessment,ayodeji2020new}.
While these approaches excel as they automate event identification, they come with two inherent drawbacks:
1) Substantial amounts of naturally scarce historical attack and fault samples are required.
2) Inability to provide information on unknown event types.
These shortcomings motivate the following question:
\textit{How can operators of \glspl{CPS} be provided with relevant information for real-time incident response, given the lack of historical critical event observations and the variety of known and unknown attack and fault types potentially affecting different physical or digital components of a \gls{CPS}?}
To address this question, this work proposes CyPhERS, a new \textbf{Cy}ber-\textbf{Ph}ysical \textbf{E}vent \textbf{R}easoning \textbf{S}ystem (see Fig.~\ref{fig:CyPhERS}).
CyPhERS comprises a two-stage process which infers event information such as occurrence, location, root cause, and physical impact from joint evaluation of network traffic and physical process data in real time. 
Stage~1 creates informative event signatures of unknown and known cyber attacks and faults by combining methods including cyber-physical data fusion, unsupervised multivariate time series anomaly detection, and anomaly type differentiation.
In Stage~2, the event signatures are evaluated either by automated matching with a signature database of known events, or through manual interpretation by the operator. 
\renewcommand{\thefigure}{1}
\begin{figure}[h]
\centering\input{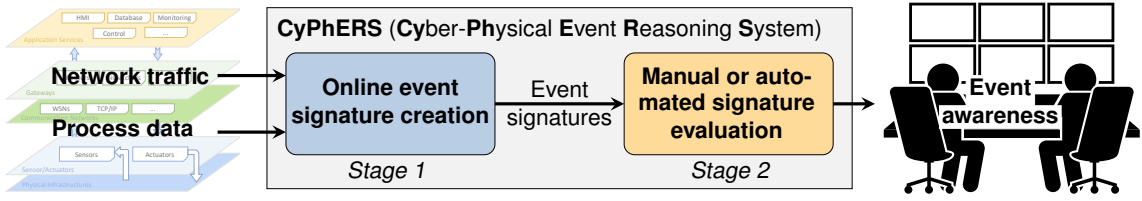}
    \caption{Schematic representation of CyPhERS.} \label{fig:CyPhERS}
\end{figure}

\subsection{Related works} \label{ssec:related_works}
Literature on attack or fault identification for \glspl{CPS} is rich \cite{9536650,Luo2021,Giraldo2018,8975131,ALGULIYEV2018212,LINDEMANN2021103498,DALZOCHIO2020103298}.
Many works propose methods which are independent of historical event observations, and able to detect both known and unknown events. 
Table~\ref{tab:concepts_comparison} summarizes these approaches and classifies them by the groups A-C.
Conceptual differences exist in the considered data sources (process data, network traffic or both), and the information they provide (e.g., event occurrence, location or impact).
A description of methods falling under group A-C is provided in Section~\ref{sssec:existing_concepts}.
Thereafter, a comparison with CyPhERS follows in Section~\ref{sssec:comparison_to_cyphers}.

\subsubsection{Existing event identification concepts for \glspl{CPS}} \label{sssec:existing_concepts}
\paragraph{Group A}
Works summarized in group A are conceptually characterized by the joint evaluation of multiple \gls{CPS} variables (e.g., multiple sensor readings), where considered variables are derived either exclusively from physical process \cite{9708436,10.1007/978-3-030-30490-4_56,10.1145/3447548.3467137} or network traffic data \cite{HUONG2021103509}.
Unsupervised multivariate time series anomaly detection is applied to detect deviations from normal behavior induced by attacks or faults.
Their output is a binary description of the system state (\textit{normal} vs. \textit{abnormal}). 
Differences among the works mainly exist in the applied \gls{ML} models and detection rules. 
The proposed methods can detect unknown and known attack or fault types in real time, given that these entail anomalies in the monitored data.
However, as they provide system-wide anomaly flags they only inform about occurrence of an event, while further information, e.g., affected devices or event type, is neglected.  
Moreover, their restriction to either network or process data limits the events they can detect.
While exclusively monitoring network data is blind to physical faults and physical impacts of cyber attacks, limiting to process data misses pure cyber events and only detects cyber-physical attacks when their impact on the system already happened.

\paragraph{Group B}
Methods falling under group B apply unsupervised multivariate time series anomaly detection either on process \cite{10.1609/aaai.v33i01.33011409,Zhang2019,tuli2022tranad, khoshnevisan2019rsm,Hallac2017,Song2018,spacecraft_AD} or network data \cite{Su2019,9355537}, similar to group A. 
Central difference is the provision of feature-level rather than system-wide anomaly flags. 
By identifying the system variables with the highest individual anomaly scores, affected components are localized in the \glspl{CPS}.
Thus, these concepts provide additional information about detected events to operators. 
However, the scores only describe the intensity of the deviation from normal behavior, while further characteristics of an anomaly (i.e., polytypic anomaly flags), for example the direction of the deviation or occurrence of missing data, are not considered.
Consequently, information on the physical impact or root causes is limited. 
Moreover, as for group A, monitoring is limited to either network or process data.
Consequently, they cannot detect and distinguish anomalies induced by both cyber attacks and physical failures.

\paragraph{Group C}
Methods in group C apply unsupervised multivariate time series anomaly detection to features of both process and network data \cite{8791598,7828584,9277640}.
Consequently, they are equally capable of detecting anomalies caused by cyber attacks and physical failures.
Moreover, compared to the subset of methods from group A and B which only monitors process data, cyber-physical attacks can be detected earlier and potentially before impacting the process.
However, only system-wide monotypic anomaly flags are provided. Therefore, these methods only inform operators about the occurrence of an event without further context information.
In contrast to the concepts represented by group A and B, literature on cyber-physical unsupervised anomaly detection for \glspl{CPS} is rare. 

\setlength{\tabcolsep}{3pt}
\begin{table}[h]
\renewcommand*{\arraystretch}{1.2}
\fontsize{8}{9}\selectfont
\caption{Comparison of CyPhERS to existing event sample-independent attack or fault identification concepts.}
\begin{center}
\begin{tabular}{c l c c c c}
\hline 
\multirow{1}{*}{\shortstack[l]{\textbf{Group}}}&\multirow{1}{*}{\shortstack[l]{\textbf{Concept description}}}&\multirow{1}{*}{\shortstack[l]{\textbf{Event} \textbf{coverage}}}&\multirow{1}{*}{\shortstack[l]{\textbf{Early}  \textbf{detection}}}&\multirow{1}{*}{\shortstack[l]{\textbf{Locali}\textbf{za\smash{ti}on}}}&\multirow{1}{*}{\shortstack[l]{\textbf{Cause \& impact} \textbf{identification}}}\\
\hline
\multirow{2}{*}{A} &\multirow{2}{*}{\shortstack[l]{Multivariate physical \textit{or} network features,\\ system-wide monotypic anomaly flags}} & \multirow{2}{*}{\shortstack[c]{o}} & \multirow{2}{*}{\shortstack[c]{o}} & \multirow{2}{*}{\shortstack[c]{\textbf{\textminus}}} & \multirow{2}{*}{\shortstack[c]{\textbf{\textminus}}}\\
& & & & &\\
\multirow{2}{*}{B} &\multirow{2}{*}{\shortstack[l]{Multivariate physical \textit{or} network features,\\ feature-level monotypic anomaly flags}} & \multirow{2}{*}{\shortstack[c]{o}} & \multirow{2}{*}{\shortstack[c]{o}} & \multirow{2}{*}{\shortstack[c]{o}} & \multirow{2}{*}{\shortstack[c]{\textbf{\textminus}}}\\
& & & & &\\
\multirow{2}{*}{C} &\multirow{2}{*}{\shortstack[l]{Multivariate physical \textit{and} network features,\\ system-wide monotypic anomaly flags}} & \multirow{2}{*}{\shortstack[c]{\textbf{+}}} & \multirow{2}{*}{\shortstack[c]{\textbf{+}}} & \multirow{2}{*}{\shortstack[c]{\textbf{\textminus}}} & \multirow{2}{*}{\shortstack[c]{\textbf{\textminus}}}\\
& & & & &\\
\multirow{2}{*}{CyPhERS} &\multirow{2}{*}{\shortstack[l]{Multivariate physical \textit{and} network features,\\ feature-level polytypic anomaly flags}} & \multirow{2}{*}{\shortstack[c]{\textbf{+}}} & \multirow{2}{*}{\shortstack[c]{\textbf{+}}} & \multirow{2}{*}{\shortstack[c]{\textbf{+}}} & \multirow{2}{*}{\shortstack[c]{\textbf{+}}}\\
& & & & &\\
\hline
\end{tabular}
\label{tab:concepts_comparison}
\end{center}
\end{table}
\renewcommand*{\arraystretch}{1}

\subsubsection{Comparison of existing concepts to CyPhERS} \label{sssec:comparison_to_cyphers}
CyPhERS combines strategies of the described concepts (group A-C), such as fusion of process and network data, unsupervised multivariate time series anomaly detection, and provision of feature-level anomaly flags (see Table~\ref{tab:concepts_comparison}). 
It further leverages their associated advantages by considering polytypic anomaly flags.
Together, this allows CyPhERS to generate highly informative and recognizable event signatures in Stage~1 (see Fig.~\ref{fig:CyPhERS}).
Moreover, CyPhERS comprises strategies for manual and automated evaluation of the event signatures (Stage~2).



\subsection{Contribution and paper structure}
The main contributions of this work are as follows:
\begin{itemize}
    \item Introduction of CyPhERS, a cyber-physical event reasoning system which provides real-time information about unknown and known attack and fault types in \glspl{CPS}, while being independent of historical event observations. 
    \item Concept demonstration, evaluation and benchmarking on a \gls{CPS} study case, considering a variety of attack and fault types affecting several system components.
    \item Discussion of possible modifications to further improve and extend CyPhERS.
\end{itemize}

The remainder of the paper is structured as follows: 
In Section~\ref{sec:cyphers_introduction}, CyPhERS is conceptually introduced.
The considered demonstration case is presented in Section~\ref{sec:demonstration_case}.
Section~\ref{sec:methodology} explains methodological details of CyPhERS, and demonstrates its implementation on the given case.
In Section~\ref{sec:results}, results of applying CyPhERS on the study case are presented.
Finally, demonstration results are discussed in Section~\ref{sec:discussion}, followed by a conclusion and view on future work in Section~\ref{sec:conclusion}.
\section{Conceptual introduction of CyPhERS} \label{sec:cyphers_introduction}
This section introduces the two stages of CyPhERS at a conceptual level.
Section~\ref{ssec:Stage_1} provides details on the online event signature creation (Stage~1).
Thereafter, signature evaluation (Stage~2) is explained in Section~\ref{ssec:Stage_2}.
Finally, Section~\ref{ssec:taxonomy} provides a taxonomy of \glspl{CPS} that CyPhERS can be applied to. 

\subsection{Online event signature creation (Stage~1)} \label{ssec:Stage_1}
Stage~1 of CyPhERS is schematically outlined in Fig.~\ref{fig:Stage_1}.
To provide event signatures that contain information about occurrence, location, root cause and physical impact of unknown and known types of both attacks and faults in real time, CyPhERS combines several strategies, which are introduced in the following.
\renewcommand{\thefigure}{2}
\begin{figure}[h]
\centering
    \input{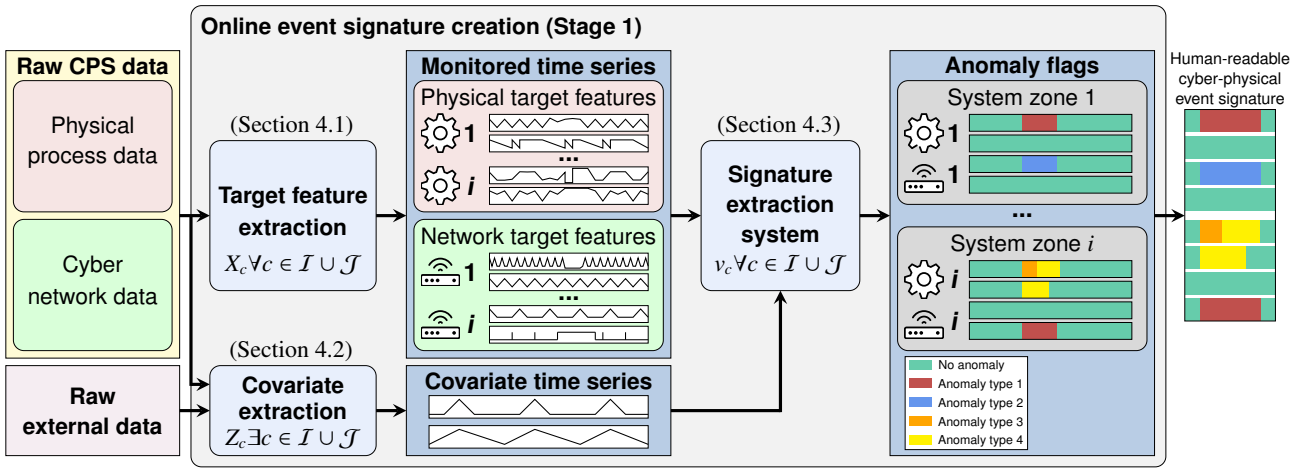}
    \caption{Schematic overview of CyPhERS' online event signature creation (Stage~1).} \label{fig:Stage_1}
\end{figure}

\subsubsection{Multi-domain information}
First element is the fusion and joint evaluation of physical process and cyber network data of a \gls{CPS} (see Fig.~\ref{fig:Stage_1}), which is required for real-time detection and differentiation of attacks and failures.  
Moreover, cyber-physical monitoring allows to determine whether a cyber attack already entails physical impact or detect it early enough in network traffic to mitigate damage through countermeasures such as isolation of affected devices. 
Other data such as maintenance activities and schedules (human domain) can potentially be included to take human errors into consideration. 

\subsubsection{Feature-level monitoring}
Second element is the individual monitoring of multiple system variables (see Fig.~\ref{fig:Stage_1}), aiming for a more detailed picture of critical events in contrast to monitoring the system state. 
Within CyPhERS, the set of monitored variables is multivariate in two dimensions:
\begin{itemize}
    \item Monitoring variables of multiple physical and network components (\textit{cross-device multivariate monitoring}).
    \item Monitoring several variables of a single component
    (\textit{in-device multivariate monitoring}). 
\end{itemize}
While cross-device multivariate monitoring aims at localizing affected devices, in-device multivariate monitoring is supposed to provide further details about a component's abnormal behavior. 
The monitored system variables are derived, for example, from sensor measurements or traffic of network devices.
In the following, the resulting set of monitored system variables is referred to as \textit{target features}, with $\mathcal{I}$ being the physical target feature and $\mathcal{J}$ the network target feature subset.
The time series of an arbitrary target feature $c$ is defined as $X_c = \{x^{c}_{1},x^{c}_{2},...,x^{c}_{N} \ | \ x^{c}_{i} \in \mathbb{R} ~\forall i\}$.
Details on the extraction of target features follow in Section~\ref{sssec:feature_extraction}. 

\subsubsection{Unsupervised time series anomaly detection considering covariates} \label{sssec:TS_AD}
CyPhERS applies unsupervised time series anomaly detection\footnote{Sometimes referred to as self-supervised anomaly detection.} to identify the occurrence of critical events.
Time series models are applied to provide normal behavior references of individual target features, which are compared to actual observations for detecting abnormal system behavior. 
Central argument is the independence of historical event observations, and ability to detect both known and unknown event types, given that they entail anomalies.  
Monitoring target features in time series format allows to detect deviations from normal behavior which are only abnormal in a specific temporal context (local anomalies) \cite{8926446}. 
Furthermore, covariates are considered for time series modeling.
Covariates allow to provide models with further system internal or external information (see Fig.~\ref{fig:Stage_1}).
As a result, situational anomalies can be detected which are only abnormal in the context of the provided covariates. 
A covariate time series of a target feature $c$ is defined as $Z_c = \{z^{c}_{1},z^{c}_{2},...,z^{c}_{N} \ | \ z^{c}_{i} \in \mathbb{R} ~\forall i\}$.
Covariate extraction is detailed in Section~\ref{sssec:covariate_extraction}.

\subsubsection{Differentiation of anomaly types}
The fourth element is the semantic differentiation of multiple anomaly types (see Fig.~\ref{fig:Stage_1}).
In case that an anomaly is flagged for a target feature $c$, it is further classified based on information such as the direction of the deviation (e.g., abnormally \textit{many} data packets received by network device X). 
Considering various anomaly types provides additional information for identification of event root causes and impact.
The series of flags provided by the signature extraction system for a target feature $c$ is given as $v_c = \{v^{c}_{1},v^{c}_{2},...,v^{c}_{N} \ | \ v^{c}_{i} \in \{-2,-1,0,1,2\} ~\forall i\}$.
A detailed explanation of the signature extraction system, including anomaly types, follows in Section~\ref{ssec:detection_system}. 

\subsubsection{Event signature visualization}
By covering multiple domains, system variables and anomaly types, Stage~1 of CyPhERS provides dense information about critical events in form of anomaly flag series of a set of target features.
To ease extraction of these information, the flag series are re-organized by grouping them for each system zone of a \gls{CPS} (see Fig.~\ref{fig:Stage_1}).
A system zone comprises a group of physical components and network devices which are directly related, such as a set of physically connected process units and a \gls{PLC} monitoring and controlling them. 
Due to the logical relation within a system zone, anomaly flags of different target features can be quickly related. 
As a result, Stage~1 of CyPhERS provides information-rich and human-readable event signatures. 

\subsection{Signature evaluation (Stage~2)} \label{ssec:Stage_2}
The concept of CyPhERS' Stage~2 is schematically depicted in Fig.~\ref{fig:Stage_2}.
In Stage~2, the event signatures of Stage~1 are evaluated, which can be realized through interpretation by human operators as well as by automated reasoning systems. 
The provided signatures are event specific and distinguishable. 
Thus, for known attacks or faults they can be predefined and stored in a database.
Once Stage~1 indicates occurrence of an event, the associated signature can be compared to the database. 
In case of a signature match, the stored information about event type, affected component, root cause, and/or physical impact provide the event hypothesis. 
Signature matching can either be conducted by the operator through visual comparison or an automated evaluation system.
One automation approach would be the transformation of a signature into a set of rules, e.g., 

\texttt{\textit{flagging of} (anomaly type 1 \textit{in} target feature X) \textit{and} (type 2 \textit{in} target feature Y)}  

\texttt{
\textit{indicates} (device A \textit{being targeted by} attack type B \textit{causing} physical impact C)}. 

Signatures can also be defined for unknown event types based on partial knowledge.
In this instance, they carry reduced information, e.g., 

\texttt{\textit{flagging of} (anomaly type 1 \textit{in} target feature X)} 

\texttt{\textit{indicates} (device A \textit{failure} [type unknown]  \textit{causing} physical impact B)}.

In case of unknown or undefined signatures, automated evaluation cannot infer event information.
In these situations, operators may deduce information such as affected system components or physical impact based on process expertise.
The minimum information CyPhERS provides in any event case is the occurrence of abnormal system behavior. 
\renewcommand{\thefigure}{3}
\begin{figure}[t!]
\centering
    \input{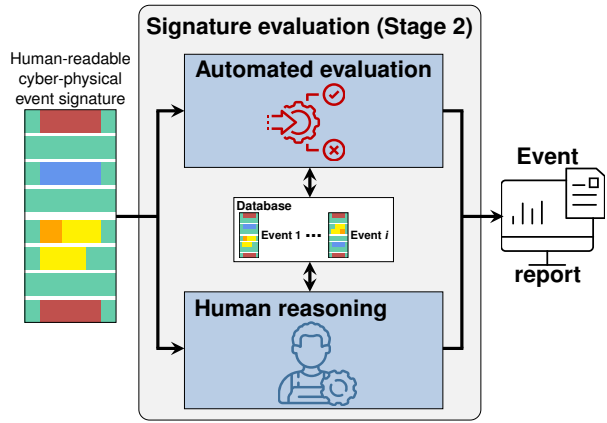}
    \caption{Schematic overview of CyPhERS' signature evaluation (Stage 2).} \label{fig:Stage_2}
\end{figure}

\subsection{Application scope of CyPhERS} \label{ssec:taxonomy}
The CyPhERS concept is applicable to \glspl{CPS} which fulfill the following requirements: 
1) Real-time data availability of and 2) learnable normal behavior patterns in both process and network traffic data.
\Gls{CPS} traffic typically exhibits periodical patterns as it arises from automated processes (e.g., polling) in static network architectures with a consistent number of devices \cite{6489745}. 
Moreover, many technical systems exhibit learnable process patterns, including manufacturing processes \cite{8953015}, transportation systems \cite{8751374}, water distribution systems \cite{abokifa2019real}, and spacecraft \cite{9405665}.
A potentially complicating factor is process volatility and randomness, which, for example, can result from unpredictable weather or user influences.
While CyPhERS is conceptually applicable to most \gls{CPS} types, its implementation requires some case specific adaptations given the heterogeneity of processes and network architectures.
These include selection or definition of target features, anomaly types, and known event signatures. 
Among instances of the same system type (e.g., health monitoring system of a specific provider), the implementation of CyPhERS is fully transferable.

\section{Demonstration case description} \label{sec:demonstration_case}
This section describes the considered demonstration case, which is introduced by Faramondi \textit{et al.} in \cite{Faramondi2021}.
The underlying \gls{CPS} is detailed in Section~\ref{ssec:case_introduction}.
Thereafter, included attack and fault scenarios, and the associated dataset are described in Section~\ref{ssec:dataset_description}.
The demonstration case was selected as the only complete cyber-physical dataset which describes various types of both cyber attacks and physical faults affecting different system components. 
The physical process represents a typical \gls{CPS} laboratory setup.   
\renewcommand{\thefigure}{4}
\begin{figure}[b]
\centering
    \input{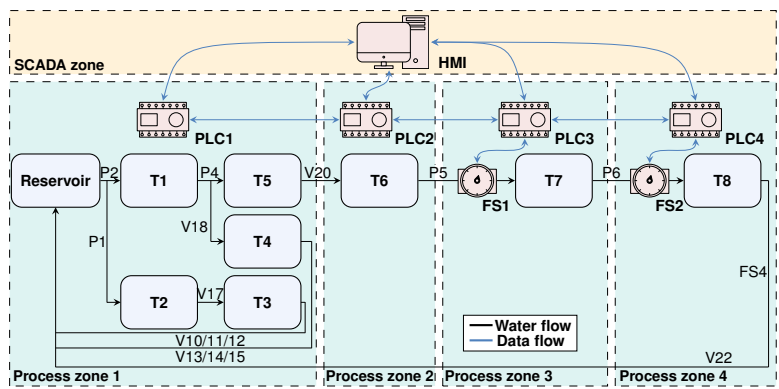}
    \caption{Simplified overview of the demonstration case \gls{CPS} based on schematic representations in \cite{Faramondi2021}.} \label{fig:Process_scheme} 
\end{figure}

\subsection{Cyber-physical structure of demonstration case} \label{ssec:case_introduction}
Fig.~\ref{fig:Process_scheme} provides a simplified overview of the investigated \gls{CPS}.
The system distributes water between several tanks, where one process cycle is defined by a filling/emptying process of all tanks. 
This procedure is continuously repeated, making it a cyclical process.
The physical system comprises eight water tanks (T1-T8), a reservoir, and several sensors and actuators.
Actuators include valves (V10-V22) and pumps (P1-P6), which realize the water distribution between the tanks. 
Note that for better readability not all sensors and actuators are depicted in Fig.~\ref{fig:Process_scheme}.
Pressure sensors in T1-T8 and \glspl{FS} (FS1-FS4) measure tank fill levels $H$ and water flows $F$, respectively.
The process is monitored and controlled by a typical \gls{SCADA} architecture consisting of the sensors and actuators (field instrumentation control layer), four \glspl{PLC} (process control layer), and a \gls{SCADA} workstation, including a \gls{HMI} and data historian (supervisory control layer).
The \gls{SCADA} workstation, \gls{HMI} and data historian together are referred to \gls{HMI} in the following. 
The communication is conducted via MODBUS TCP/IP protocol.
The process consists of four stages, each of which is controlled by one of the four \glspl{PLC}. 
The \glspl{PLC} send sensor values to the \gls{SCADA} workstation, so that physical process data can be stored centrally on the historian. 
Moreover, values of tank fill levels $H$ and water flows $F$ are directly exchanged between the \glspl{PLC} and \glspl{FS}, which they require to control tank fill levels by (de-)activating pumps and valves. 
While most sensors and actuators are connected to the \glspl{PLC} via wired links,  \gls{FS}1 and \gls{FS}2 are MODBUS TCP/IP sensors with own IP addresses. 
Thus, the communication network in total consists of seven devices, which are marked red in Fig.~\ref{fig:Process_scheme}.
Note that an additional Kali Linux machine was used to launch cyber attacks, which is not depicted in Fig.~\ref{fig:Process_scheme}.

\subsection{Threat scenarios and dataset} \label{ssec:dataset_description}
The dataset comprises four partitions, each covering multiple process cycles.
While the first partition describes a normal operation scenario (S0) the remaining three describe attack and fault scenarios (S1-S3).
S1-S3 exhibit an increasing level of event type variety.  
S1 includes several physical component breakdowns and water leaks as well as \gls{MITM} attacks.
In \gls{MITM} attacks a perpetrator positions himself between two victim devices to relay and potentially alter communication while the victims assume a direct communication \cite{7442758}. 
In the present case, the attacker modifies $H$ values send by victim one, which are required by victim two to control fill levels of tanks in the respective process zone. 
In S2, \gls{DoS} attacks are additionally included.
These cause a disconnection of the targeted device from the network by flooding it with requests \cite{Mahjabin2017}.
In the investigated dataset, several \gls{DoS} attack variants are used to disconnect specific \glspl{PLC} or the \gls{SCADA} workstation.
Finally, S3 adds scanning attacks.
Scanning is a reconnaissance method used by attackers to determine possible vulnerabilities by searching for services and service identifiers in a target network or host \cite{6657498}.
The given case considers several scanning attacks, which are used to gather information about various \glspl{PLC}.
Note that S1-S3 comprise of unique attack and fault events affecting various components and communications links in the system, so that each scenario represents an entirely new case. 
In total, eight \gls{MITM}, five \gls{DoS}, and seven scanning attacks as well as three water leaks and six sensor or pump breakdowns are included.
The considered attack types are among the most relevant for \glspl{CPS} \cite{HASAN2023103540, 9019636, 9225126,YAACOUB2020103201}. 
Table~\ref{tab:features} lists the raw physical and network features of the dataset.
The physical data has a constant one-second resolution, resulting in 3429 (S0), 2421 (S1), 2105 (S2), and 1255 (S3) samples. 
The network data during normal operation (S0) on average contains 2265 packets per second, and in total comprises $\sim$7.8 (S0), $\sim$5.5 (S1), $\sim$5.2 (S2), and $\sim$5.9 (S3) $\times10^{6}$ packets. 
For a more detailed explanation of the demonstration case the reader is referred to \cite{Faramondi2021}.
\begin{table}[h]
\fontsize{8}{9}\selectfont
\caption{Raw network and process features within the study case.}
\begin{center}
\begin{tabular}{c l c l}
\hline
\textbf{No.}&\textbf{Physical features}&\textbf{No.}&\textbf{Network features}\\
\hline
$1$ & Timestamp & $1$ & Timestamp \\

\multirow{1}{*}{$2$-$9$} & \multirow{1}{*}{\shortstack[l]{Fill level $H$ of T1-T8}} & $2$-$3$ & IP address (src. \& dst.)$^{\mathrm{a}}$\\

$10$-$15$ & \begin{tabular}{@{}l@{}}Activation state $S$ of P1-P6\end{tabular} & $4$-$5$ & MAC address (src. \& dst.)\\

\multirow{2}{*}{$16$-$19$} & \multirow{2}{*}{\shortstack[l]{Flow value $F$ measured by \\ \gls{FS}1-\gls{FS}4}} & $6$-$7$ & Port (src. \& dst.)\\

&   & $8$ & Protocol\\

$20$-$41$ & Activation state $S$ of V1-V22  & $9$ & TCP flags \\

 & & $10$ & Packet size \\

 & & $11$ & MODBUS function code \\

 & & $12$ & MODBUS response value \\
 
 & & $13$-$14$ & No. of packets (src. \& dst.) \\
\hline
\multicolumn{4}{l}{$^{\mathrm{a}}$Src. and dst. refer to source and destination, respectively.} \\
\end{tabular}
\label{tab:features}
\end{center}
\end{table}

\section{Methodology and implementation of CyPhERS} \label{sec:methodology}
This section first provides details on the methodology and case-specific implementation of the online event signature creation (Stage~1).
This includes extraction of target features (Section~\ref{sssec:feature_extraction}) and covariates (Section~\ref{sssec:covariate_extraction}) as well as the signature extraction system (Section~\ref{ssec:detection_system}).
Thereafter, methodology and case-specific implementation of the event signature evaluation (Stage~2) is detailed in Section~\ref{sssec:manual_reasoning}.

\subsection{Target feature extraction} \label{sssec:feature_extraction}

\paragraph{Methodology}
The landscape of \glspl{CPS} is characterized by a pronounced heterogeneity, resulting from factors such as the diversity of physical components (e.g., tanks, engines or batteries) and communication protocols (e.g., MODBUS, UDP or DNP). 
Thus, a general set of target features cannot be defined. 
Nevertheless, guidelines for extraction of relevant features can be provided.


CyPhERS considers sensor measurements and actuator states as raw physical process data.
Monitoring such data has two motivations, namely the identification of i) true physical events and ii) manipulation of process-relevant data.
The former requires features which represent the behavior (e.g., state or output) of all physical components of a \gls{CPS} to 1) localize affected devices and 2) infer the impact on them.
For the latter, readings exchanged among devices for automated process control need to be monitored. 
For processes exhibiting low randomness and noise levels, raw sensor readings can directly be used as physical target features. 
Otherwise, further processing is required to extract the available information. 
Strategies include resampling (e.g., moving average) or derivation of  features which describe a component's behavior on a simplified level (e.g., on/off state).

In CyPhERS, network target features are extracted from OT network traffic\footnote{Other potential data sources include system logs and key performance indicators of digital devices (e.g., memory usage).} of a \gls{CPS}.
Traffic monitoring is considered to retain information for i) localizing affected digital devices, and ii) concluding on attack types. 
For the former, traffic of each network device is monitored separately. 
The latter requires extraction of several features for each device in order to provide sufficient information for distinguishing attack types. 
Detecting and differentiating attacks is challenged by the fact that some solely concern individual packets (e.g., sending a malicious control command), while others are only visible from the context of multiple packets (e.g., replaying valid data transmission). 
Therefore, CyPhERS considers both extraction of network features which i) are sensible to values of single packets (e.g., count of packets send from unknown source IP or MAC addresses), and ii) set multiple packets into context (e.g., average number of received packets within a time period). 


\paragraph{Case-specific implementation} 
In the present demonstration case, noise in the raw physical features is  comparatively low, which allows direct use as target features with original per-second resolution.
For the sake of clarity, not all available sensor readings and actuators states are taken into account.
Instead, fill levels $H$ of the water tanks are considered, as they allow to monitor all four process steps, and describe the behavior of the most important components.
The network traffic is separately monitored for \gls{PLC}1-4, FS1-2 and the \gls{HMI}. 
For this purpose, it is first filtered by the destination MAC addresses and then evaluated for each second through several features.
An overview of the resulting set of physical $\mathcal{I}$ and network target features $\mathcal{J}$ can be found in Table~\ref{tab:target_features}.
\setlength{\tabcolsep}{1.7pt}
\begin{table}[h]
\fontsize{8}{9}\selectfont
\caption{Overview of target features extracted in the study case.} 
\begin{center}
\begin{tabular}{c l c l}
\hline
\textbf{No.}&\textbf{Physical target features \bm{$\mathcal{I}$}}&\textbf{No.}&\textbf{Network target features$^{\mathrm{a}}$ \bm{$\mathcal{J}$}}\\
\hline
$1$-$8$ & Fill level $H$ of T1-T8 & $1$-$7$ & Average packet size $s_{\text{packet}}$\\

& & $8$-$14$ & Packet count $n_{\text{packet}}$\\

&  & $15$-$21$ & New src. IP/MAC count $n_{\text{IP/MAC}}$\\

&  & $22$-$28$ & New TCP flags count $n_{\text{TCP}}$\\

&  & $29$-$35$ & Mean of encoded TCP flags$^{\mathrm{b}}$ $\mu_{\text{TCP}}$\\

&  & $36$-$42$ & Different src. ports count $n_{\text{ports}}$\\

\hline
\multicolumn{4}{l}{$^{\mathrm{a}}$For \gls{PLC}1-4, FS1-2 and \gls{HMI}, each considered as destination device.} \\
\multicolumn{4}{l}{$^{\mathrm{b}}$Each flag type is encoded as a specific integer.}
\end{tabular}
\label{tab:target_features}
\end{center}
\end{table}


\renewcommand{\thefigure}{5}
\begin{figure}[b]
\centering
\input{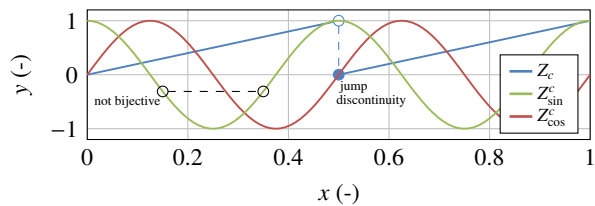}
\caption{Illustration of sine and cosine transformation of cyclical covariates.} \label{fig:sine_cosine}
\end{figure}
\subsection{Covariate extraction} \label{sssec:covariate_extraction}
\paragraph{Methodology}
As for target features, extraction of covariates is dependent on the respective \gls{CPS}. 
In any case, however, they should represent process-relevant context such as environmental conditions or human interactions.
Examples are irradiation for solar plants or user intervention in case of self-driving cars. 
Such information allows models to learn whether an event is normal or abnormal given the current context.
For example, irradiation facilitates differentiation of normal weather- or malicious attacker-induced drops of solar feed-in power.
\Glspl{CPS} often exhibit repeating processes.
In such cases, covariates which inform models about the current position in a cycle (e.g., process stage or time of the day) provide valuable context information.
CyPhERS considers sine and cosine transformation \cite{Chakraborty} of such cyclical covariates, which is schematically represented in Fig.~\ref{fig:sine_cosine}.
Let $Z_c = \{z^{c}_{1},z^{c}_{2},...,z^{c}_{N} \ | \ z^{c}_{i} \in \mathbb{R} ~\forall i\}$ be a cyclical covariate time series of length $N$ for a target feature $c$.
Within one cycle, values of $Z_c$ are linearly increasing on the range $z^{c}_{i} \in [\min (Z_c), \max (Z_c)]$.
Due to the jump discontinuity between two cycles, a linear representation cannot properly describe the continuity of cyclical processes. 
To eliminate the discontinuity, values of $Z_c$ are transformed according to 
\begin{equation}
      z^{c}_{\text{sin},i} = \sin\left(\frac{2\pi z^{c}_{i}}{\max (Z_c)}\right),  \ \text{and} \  z^{c}_{\text{cos},i} = \cos\left(\frac{2\pi z^{c}_{i}}{\max (Z_c)}\right),  \label{eq:sine_cosine_transform}
\end{equation}
$\forall i \in [1,N]$, resulting in the two new covariate time series $Z^c_{\text{sin}}$ and $Z^c_{\text{cos}}$.
The use of both sine and cosine transformation is required as they individually are not bijective, which would lead to ambiguity in the transformed covariate (see Fig.~\ref{fig:sine_cosine}).

\paragraph{Case-specific implementation} 
In the investigated study case, the normal progress of a process cycle is considered as covariate time series $P_c = \{p^{c}_{1},p^{c}_{2},...,p^{c}_{N} \ | \ p^{c}_{i} \in \mathbb{R} ~\forall i\}$ of length $N$, $\forall c \in \mathcal{I}$.
Using S0 (normal operation scenario) the usual duration $d_{c}$ of a process cycle is determined.
Based on the duration, values of $P_c$ are defined on the range $p^{c}_{i} \in [0,d_c]$.
The additional covariate time series $P^{c}_{\text{sin}}$ and $P^{c}_{\text{cos}}$ are extracted $\forall c \in \mathcal{I}$ by applying sine and cosine transformation on the values of $P_c$ according to \eqref{eq:sine_cosine_transform} with $\max (P_c) = d_c$.

\subsection{Signature extraction system} \label{ssec:detection_system}
The signature extraction system (see Fig.~\ref{fig:Stage_1}) follows the 
idea of applying individual anomaly detection and classification pipelines to each target feature. 
Compared to the joint processing in one large model, several advantages exist: 1) The complexity of time series models can be adjusted to individual target features.
As features in CyPhERS originate from very different sources (process data and network traffic), they exhibit strong variations in characteristics such as observation rates and noise levels.
2) Independent definition of abnormal behavior for each system component. 
By selecting covariates for individual target features, it is possible to define which context models should consider when deciding whether a component is behaving abnormally. 
3) Promotes a distributed implementation of CyPhERS on edge devices.
As attackers can manipulate data to hide induced physical impact from centralized monitoring \cite{Giraldo2018}, this facilitates detection of hidden process manipulations. 

The anomaly detection and classification pipelines are explained in Section~\ref{sssec:detection_pipeline}.
After that, Section~\ref{ssec:forecasting_models} addresses the forecasting models which are applied within the pipelines. 
Finally, the procedure for automated implementation of the signature extraction system is detailed in Section~\ref{sssec:modeling_procedure}.

\subsubsection{Anomaly detection and classification pipelines} \label{sssec:detection_pipeline}
\paragraph{Methodology}
The anomaly detection and classification pipeline of a target feature $c$ is schematically depicted in Fig.~\ref{fig:Anomaly_detector}.
\renewcommand{\thefigure}{6}
\begin{figure}[b]
\centering
    \input{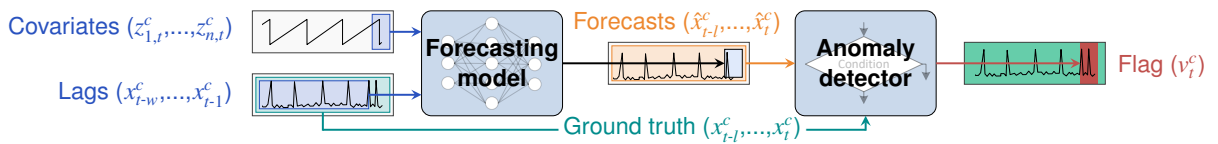}
    \caption{Schematic overview of the anomaly detection and classification pipeline of a target feature $c$.} \label{fig:Anomaly_detector}
\end{figure}
A pipeline consists of two fundamental and consecutive steps, namely a time-series forecasting model and an anomaly detector. 
Given $X_c = \{x^{c}_{1},x^{c}_{2},...,x^{c}_{N} \ | \ x^{c}_{i} \in \mathbb{R} ~\forall i\}$ and $Z_{c,1},...,Z_{c,n} = \{\{z^{c}_{1,1},z^{c}_{1,2},...,z^{c}_{1,N}\},..., \{z^{c}_{n,1},z^{c}_{n,2},...,z^{c}_{n,N}\}  |  z^{c}_{j,i}\thickmuskip=3mu\in\mathbb{R} \forall (j,i)\}$ of a target feature $c$, the forecasting model predicts the expected value $\hat{x}^{c}_{t}$ at time $t$ based on lag values $x^{c}_{t-w},...,x^{c}_{t-1}$ and covariates $z^{c}_{1,t},...,z^{c}_{n,t}$ according to
\begin{equation}
    \hat{x}^{c}_t = \Phi\left([x^{c}_{t-w},...,x^{c}_{t-1}], [z^{c}_{1,t},...,z^{c}_{n,t}]\right), \label{eq:mapping_function}
\end{equation}
where $w$ is the length of the history window and $n$ the number of covariates.
Depending on the target feature, $x^{c}_{t-w},...,x^{c}_{t-1}$ and $z^{c}_{1,t},...,z^{c}_{n,t}$ are only partially used as model input, which is specified in Section~\ref{ssec:forecasting_models}.

Next, the expected value $\hat{x}^{c}_{t}$ and ground truth ${x}^{c}_{t}$ are forwarded to the anomaly detector.
In CyPhERS, anomalies are flagged based on multiple consecutive observations instead of only the most recent one, aiming at reducing noise-induced \glspl{FP}. 
For that purpose, the detector first calculates the average of the distances of the last $l$ observations to their respective expected values according to
\begin{equation}
    \varepsilon^c_{t} = \frac{\sum^{l-1}_{j=0} |x^c_{t-j}-\hat{x}^c_{t-j}|}{l}.   \label{eq:aggregated_distance}
\end{equation}
Based on $\varepsilon^c_{t}$ and further characteristics of the current target feature observations, the detector then differentiated several anomaly types. 
While the definition of meaningful anomaly types is facilitated by taking process specificities of a \gls{CPS} into account, some widely applicable ones exist.
Table~\ref{tab:flag_description} defines some of them.
\renewcommand*{\arraystretch}{0.9}
\begin{table}[t]
\fontsize{8}{9}\selectfont
\caption{Description of some general anomaly types.}
\begin{center}
\begin{tabular}{c l l l}
\hline
\textbf{Flag $\bm{v}$}&\textbf{Anomaly type}&\textbf{Description} & \textbf{Schematic}\\
\hline
\multirow{4}{*}{\phantom{$-$}$2$} & \multirow{4}{*}{\shortstack[l]{Positive\\ disrupted}} & \multirow{4}{*}{\shortstack[l]{Target feature positively differen-\\tiates from normal behavior, ex-\\hibiting static or NaN values.}} & \multirow{4}{*}{\includegraphics[clip, trim=1.5cm 1.5cm 1.5cm 1.5cm,width=0.075\textwidth, height=10mm]{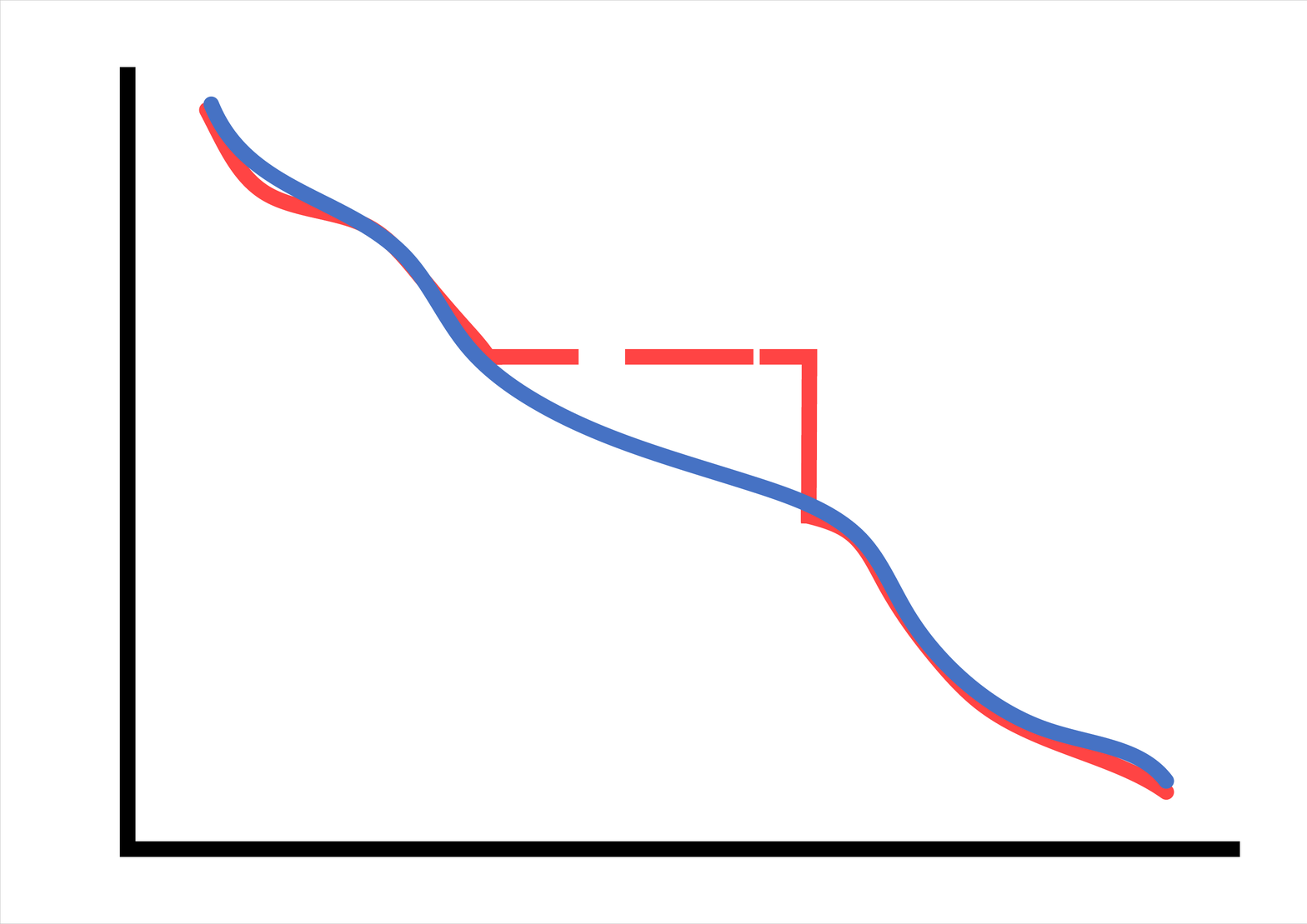}} \\
 &  &  &\\
  &  &  &\\
    &  &  &\\
\multirow{4}{*}{\phantom{$-$}$1$} & \multirow{4}{*}{\shortstack[l]{Positive\\ undisrupted}} & \multirow{4}{*}{\shortstack[l]{Target feature positively differen-\\tiates from normal behavior, not\\ exhibiting static or NaN values.}} &\multirow{4}{*}{\raisebox{-0cm}{\includegraphics[clip, trim=1.5cm 1.5cm 1.5cm 1.5cm,width=0.075\textwidth, height=10mm]{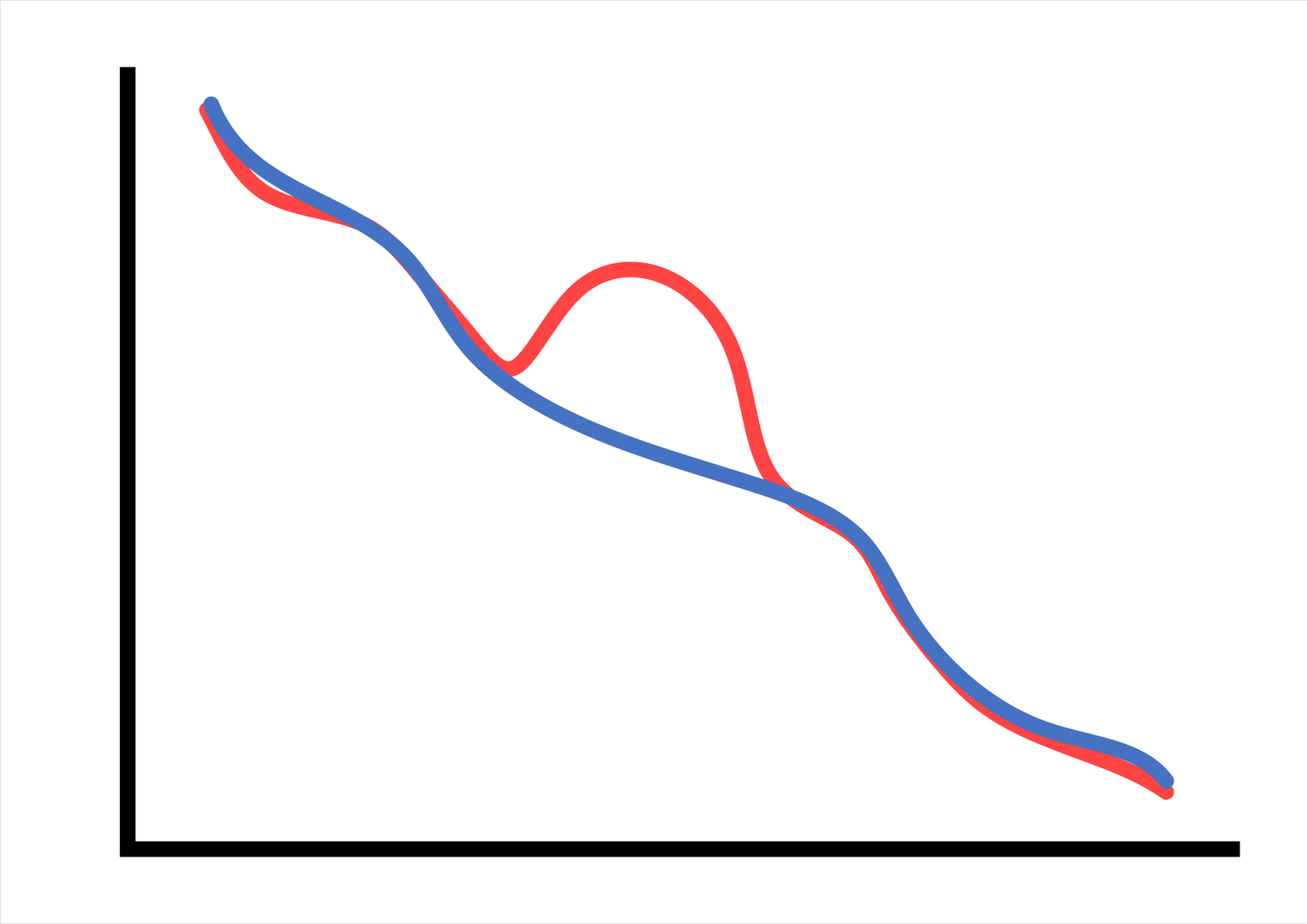}}} \\
 &  &  &\\
   &  &  &\\
  &  &  &\\
\multirow{4}{*}{$-1$} & \multirow{4}{*}{\shortstack[l]{Negative\\ undisrupted}} & \multirow{4}{*}{\shortstack[l]{Target feature negatively differen-\\tiates from normal behavior, not\\ exhibiting static or NaN values.}} & \multirow{4}{*}{\raisebox{-0cm}{\includegraphics[clip, trim=1.5cm 1.5cm 1.5cm 1.5cm,width=0.075\textwidth, height=10mm]{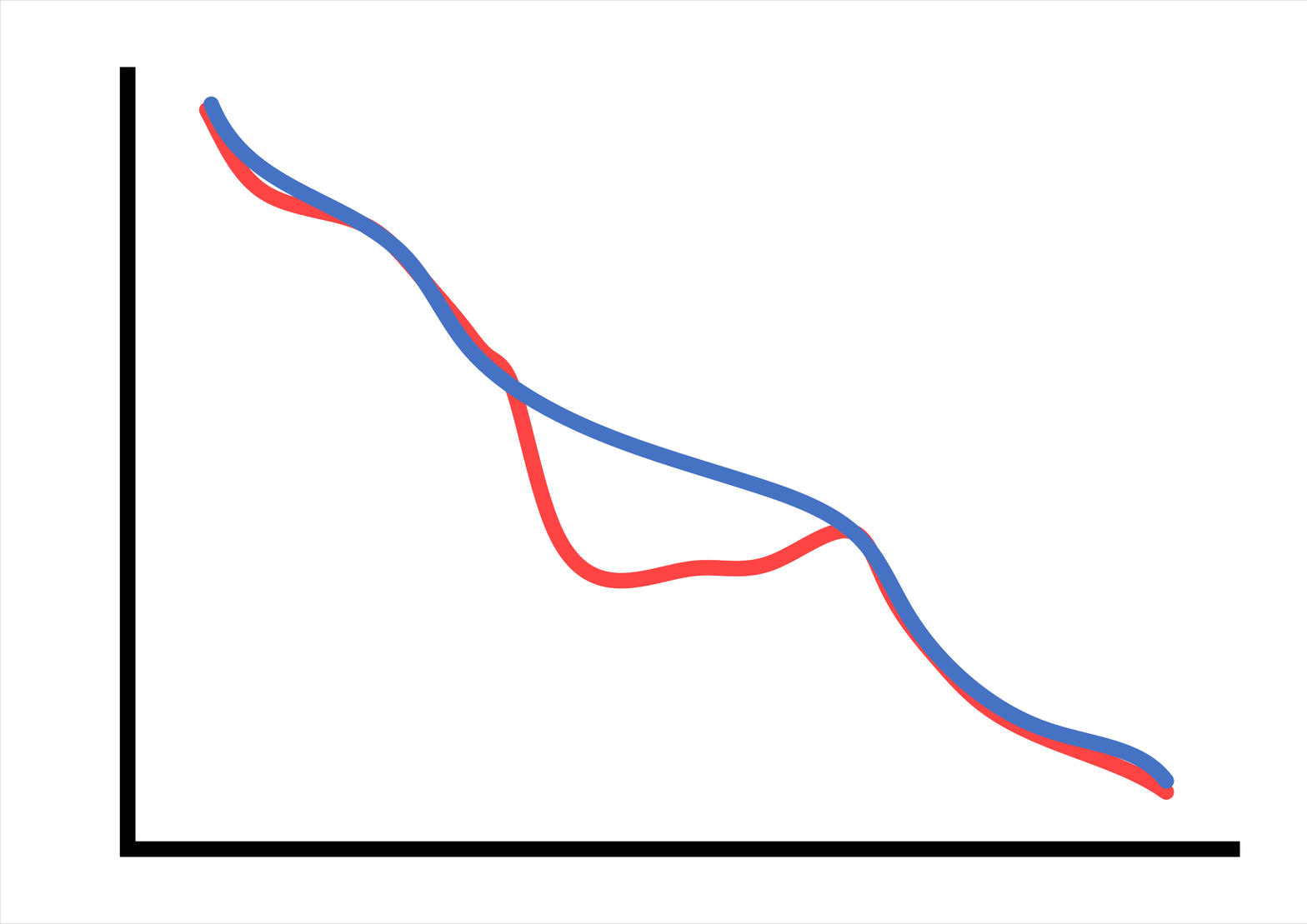}}} \\
 &  &  &\\
   &  &  &\\
  &  &  &\\
\multirow{4}{*}{$-2$} & \multirow{4}{*}{\shortstack[l]{Negative\\ disrupted}} & \multirow{4}{*}{\shortstack[l]{Target feature negatively differen-\\tiates from normal behavior, ex-\\hibiting static or NaN values.}}  &\multirow{4}{*}{\raisebox{-0cm}{\includegraphics[clip, trim=1.5cm 1.5cm 1.5cm 1.5cm,width=0.075\textwidth, height=10mm]{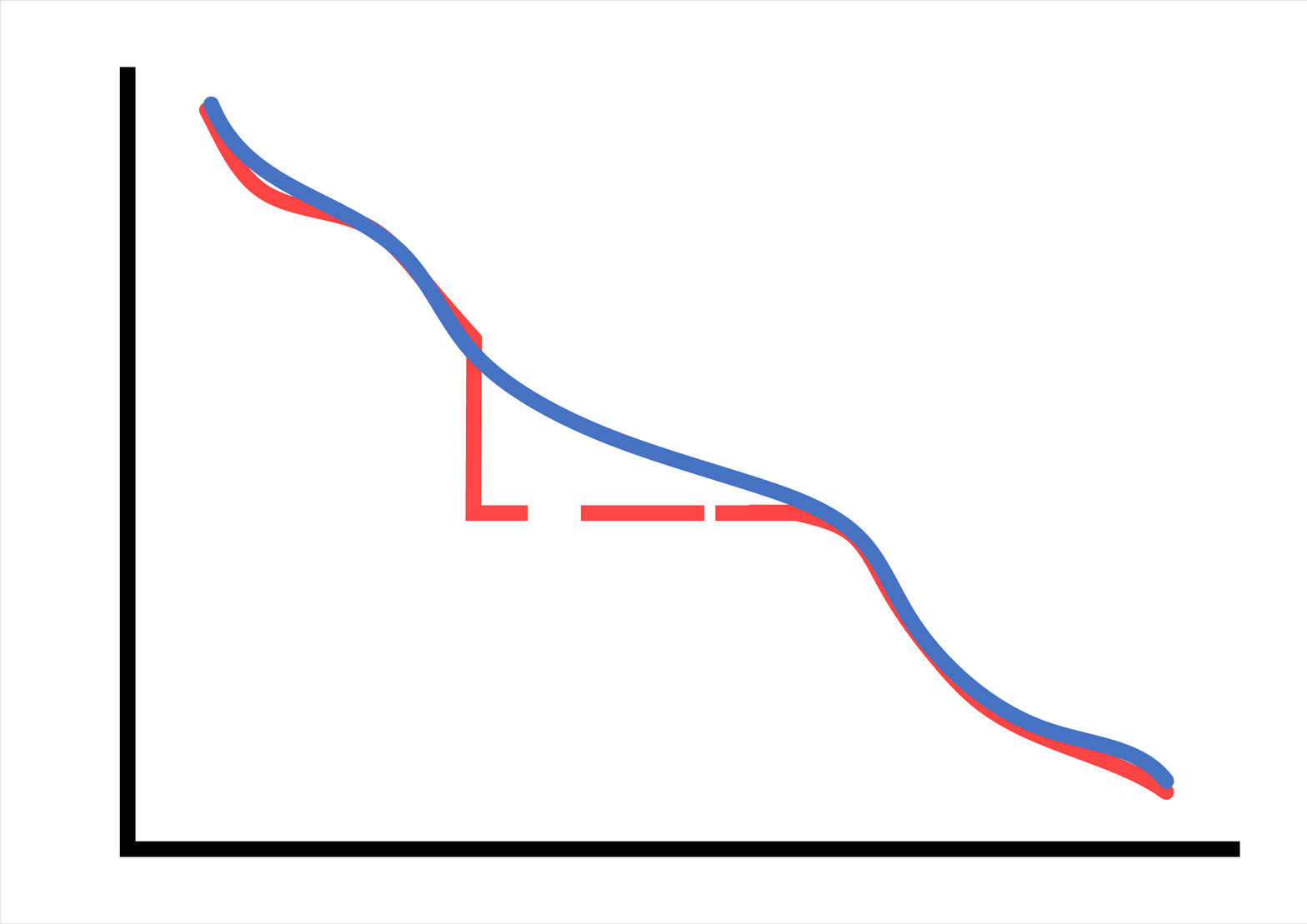}}}\\
 &  &  &\\
   &  &  &\\
  &  &  &\\
  \hdashline
\multirow{4}{*}{\phantom{$-$}$0$} & \multirow{4}{*}{No anomaly} & \multirow{4}{*}{\shortstack[l]{No abnormal behavior.}} &\multirow{4}{*}{\raisebox{-0cm}{\includegraphics[clip, trim=1.5cm 1.5cm 1.5cm 1.5cm,width=0.075\textwidth, height=10mm]{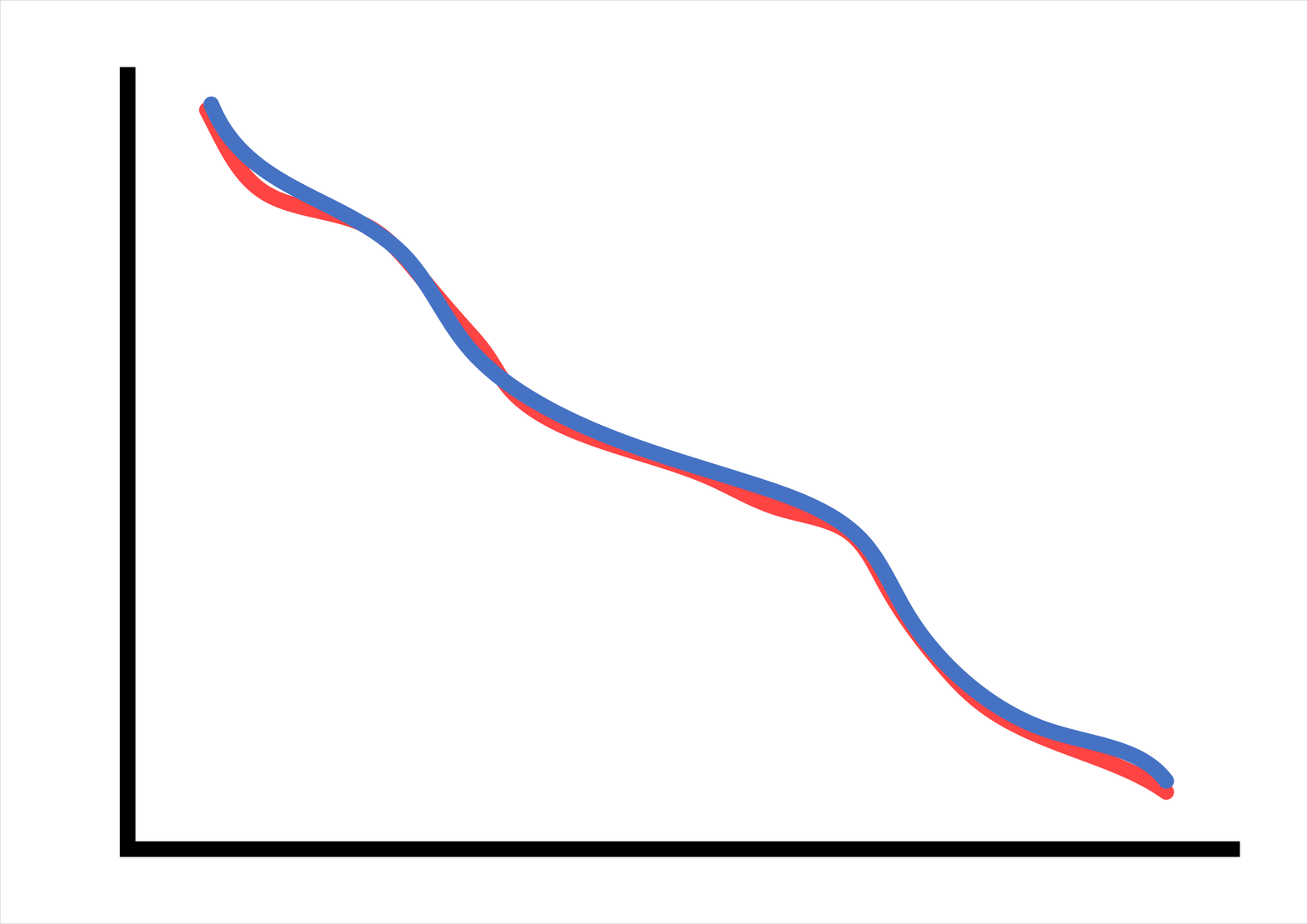}}} \\
 &  &  &\\
   &  &  &\\
   &  &  &\\
\hline
\end{tabular}
\label{tab:flag_description}
\end{center}
\end{table}
\renewcommand*{\arraystretch}{1.0}
The listed types can be transferred into the detector's anomaly flag decision function according to
\vspace{0.3cm}
\begin{equation} 
v^c_t=
\begin{cases}
    \text{\phantom{$-$}}2 & \smash{\text{if }( \overbrace{\varepsilon^c_{t}  > \tau_c}^{\text{Detection}}),\ (\overbrace{x^c_t>\Hat{x}^c_t}^{\text{Direction}}) \text{ and }[\overbrace{\exists  (x^c_{i}=x^c_{i-1}\text{ or } \text{NaN}) \in X^c_J}^{\text{Disruption}}]}\\
    \text{\phantom{$-$}}1& \text{if } (\varepsilon^c_{t}  > \tau_c),\ (x^c_t>\Hat{x}^c_t) \text{ and }[\nexists  (x^c_{i}=x^c_{i-1}\text{ or } \text{NaN}) \in X^c_J] \\
    -1& \text{if } (\varepsilon^c_{t}  > \tau_c),\ (x^c_t<\Hat{x}^c_t) \text{ and } [\nexists  (x^c_{i}=x^c_{i-1}\text{ or } \text{NaN}) \in X^c_J]\\
    -2& \text{if } (\varepsilon^c_{t}  > \tau_c),\ (x^c_t<\Hat{x}^c_t) \text{ and } [\exists  (x^c_{i}=x^c_{i-1}\text{ or } \text{NaN}) \in X^c_J] \\
    \text{\phantom{$-$}}0& \text{otherwise},
\end{cases}
\label{eq:decision_function}
\end{equation}
where $\tau_c$ is a target feature-specific threshold, and $X^c_J=x^c_{t-l+1},...,x^c_t$ the ground truth values within the distance averaging window.
The automated adaption of $\tau_c$ to individual target features is described in Section~\ref{sssec:modeling_procedure}. 
The differentiation between static and non-static behavior is neglected for target features exhibiting static values during normal operation, which in this event reduces \eqref{eq:decision_function} to
\vspace{0.5cm}
\begin{equation} 
    v^{c*}_t =   
\begin{cases}
    \text{\phantom{$-$}}2& \smash{\text{if }( \overbrace{\varepsilon^c_{t}  > \tau_c}^{\text{Detection}}),\ (\overbrace{x^c_t>\Hat{x}^c_t}^{\text{Direction}}) \text{ and }[\overbrace{\exists  (x^c_{i}=\text{NaN}) \in X^c_J}^{\text{Disruption}}]}\\
    \text{\phantom{$-$}}1& \text{if } (\varepsilon^c_{t}  > \tau_c),\ (x^c_t>\Hat{x}^c_t) \text{ and }[\nexists  (x^c_{i}=\text{NaN}) \in X^c_J] \\
    -1& \text{if } (\varepsilon^c_{t}  > \tau_c),\ (x^c_t<\Hat{x}^c_t) \text{ and } [\nexists  (x^c_{i}=\text{NaN}) \in X^c_J]\\
    -2& \text{if } (\varepsilon^c_{t}  > \tau_c),\ (x^c_t<\Hat{x}^c_t) \text{ and } [\exists  (x^c_{i}=\text{NaN}) \in X^c_J] \\
    \text{\phantom{$-$}}0& \text{otherwise}.
\end{cases} \label{eq:decision_function_2}
\end{equation}

\paragraph{Case-specific implementation} 
For the investigated study case, a distance averaging window $l=10$ is selected for calculation of $\varepsilon^c_{t}$ according to \eqref{eq:aggregated_distance}, $\forall c \in \mathcal{I}$ and $\mathcal{J}$.
Considered anomaly types correspond to the ones listed in Table~\ref{tab:flag_description}.
For $H_{\text{T}7}$ and $H_{\text{T}8}$ as well as $n_{\text{IP/MAC}}$ and $n_{\text{TCP}}$ of all network devices, the detector's decision function reduces to \eqref{eq:decision_function_2} as they contain static values during S0.

\subsubsection{Forecasting models} \label{ssec:forecasting_models}
\paragraph{Methodology}
As some \glspl{CPS} come with computational constraints, keeping model complexity at a required minimum is favorable.
The forecasting models used within the anomaly detection and classification pipelines (see Fig.~\ref{fig:Anomaly_detector}) are interchangeable, which allows adapting their complexity to the specific characteristics of a target feature.
Four target feature property classes are differentiated in CyPhERS.
These are listed in Table~\ref{tab:model_types} together with feature examples and recommended forecasting model types.
\begin{table}[b]
\fontsize{8}{9}\selectfont
\caption{Target feature property classes and proposed forecasting models.}
\begin{center}
\begin{tabular}{c l l l}
\hline
\textbf{Class}&\textbf{Condition$^{\mathrm{a}}$}&\textbf{Model type}&\textbf{Feature example}\\
\hline
A & \begin{tabular}{@{}l@{}} Target features with\\ solely constant values \end{tabular} & \begin{tabular}{@{}l@{}} Constant value \end{tabular} & \begin{tabular}{@{}l@{}} Occurrence of new\\ IP address  \end{tabular}\\
B & \begin{tabular}{@{}l@{}} Continuous target fea-\\tures with covariates \end{tabular} &\begin{tabular}{@{}l@{}} Simple regression\\ (e.g., linear or RF)\end{tabular} & Sensor measurements \\
C & \begin{tabular}{@{}l@{}} Target features with\\ discrete levels \end{tabular} & \begin{tabular}{@{}l@{}} Ensemble regre-\\ssion (e.g., RF)\end{tabular}& \begin{tabular}{@{}l@{}}States of actuators \end{tabular}\\
D & \begin{tabular}{@{}l@{}} Continuous target fea-\\tures without covariates \end{tabular} & \begin{tabular}{@{}l@{}}Deep-learning\\ (e.g., LSTM)\end{tabular} & \begin{tabular}{@{}l@{}} Number of transmitted\\network packets \end{tabular}\\
\hline
\multicolumn{4}{l}{\begin{tabular}{@{}l@{}} $^{\mathrm{a}}$Constant, continuous and discrete behavior relates to normal operation.\end{tabular}} \\
\end{tabular}
\label{tab:model_types}
\end{center}
\end{table}
Class A comprises target features exhibiting unchanged values during normal operation. 
In these cases, a trivial constant-value forecast is sufficient, which simplifies \eqref{eq:mapping_function} to $\hat{x}^c_t = a_c$, where $a_c$ corresponds to the constant value of the respective target feature during normal operation.
In class B, continuous target features with covariate availability are summarized, which typically includes physical sensor measurements. 
Due to the additional information covariates provide, less complex forecasting models can be considered.
The use of simple regression models (e.g., linear regression) for modeling class B features according to \eqref{eq:mapping_function} is recommended.
Target features may also exhibit discrete values (Class C) as in the case of actuator states.
In this event, the use of ensemble models such as a \gls{RF} regressor \cite{breiman2001random} is proposed for modeling features according to \eqref{eq:mapping_function}. 
The rationale behind using ensemble models for class C features is their internal process of discretizing continuous variables, which facilitates prediction of sudden steps. 
Moreover, they are known for robustness, few parameters to tune and good performance compared to many other standard methods on a variety of prediction problems \cite{Scornet2015,BOJER2021587}. 
For a detailed theoretical description of ensemble models, the reader is referred to \cite{hastie2009elements}.
Finally, Class D comprises continuous target features without availability of covariates. 
Since covariates are neglected, \eqref{eq:mapping_function} reduces to
\begin{equation}
    \hat{x}^{c}_t = \Phi([x^c_{t-w},...,x^c_{t-1}]). \label{eq:mapping_function_LSTM}
\end{equation}
Due to lack of additional information through covariates, more advanced models are required, which are capable of exploiting short and long-term temporal dependencies within a target feature. 
Thus, deep-learning-based forecasting models are suggested for class D features. 
Prominent representatives are \gls{LSTM} networks~\cite{6795963}.
\Glspl{LSTM} constitute a special architecture of neural networks capable of capturing complex long-term temporal dependencies in sequential data, which makes them well suited for time series forecasting. 
Many works have demonstrated their superior performance in various areas \cite{8614252,7966019,SRIVASTAVA2018232}. 
For a detailed theoretical description of \gls{LSTM} networks, the reader is referred to \cite{Goodfellow-et-al-2016}.

\paragraph{Case-specific implementation} 
In the investigated demonstration case, $n_{\text{IP/MAC}}$ and $n_{\text{TCP}}$ constitute constant target features (class A), which hence are modeled with trivial constant value forecasts.
All considered physical target features of the study case ($H_{\text{T}1}$-$H_{\text{T}8}$) fall into class B. 
Thus, \Gls{RF} is selected as the regression model.
As $P_c$, $P^c_{\text{sin}}$ and $P^c_{\text{cos}}$ are available as supporting covariate time series, a comparatively short history window of $w=10$ is considered.
Table~\ref{tab:RFR_hyperparameter} lists the tuned hyperparameters and their respective search spaces.
The underlying model selection procedure is detailed in Section~\ref{sssec:modeling_procedure}.
The \gls{RF} models are implemented in Python using the forecasting-library \textit{Darts} \cite{herzen2021darts}.

For forecasting the network target features, no supporting covariates are available. 
Thus, the set of non-constant network target features $\mathcal{J}_{\text{nc}}$ falls into property class D.
Consequently, concerned features are modeled using \gls{LSTM} networks.
To allow a \gls{LSTM} to capture long-term dependencies, the history window is extended to $w=300$, covering an entire process cycle.
In Table~\ref{tab:RFR_hyperparameter}, the tuned hyperparameters and their respective search spaces are given.
Implementation of the \gls{LSTM} models is realized in Python using the forecasting-library \textit{Darts} \cite{herzen2021darts}.

\setlength{\tabcolsep}{8.5pt}
\renewcommand{\arraystretch}{1}
\begin{table}[h!]
\fontsize{8}{9}\selectfont
\caption{Hyperparameters and search spaces for \gls{RF} and \gls{LSTM}.}
\begin{center}
\begin{tabular}{l l l}
\hline
\textbf{No.}&\textbf{Hyperparameter$^{\mathrm{a}}$}&\textbf{Search space} \\
\hline
\multicolumn{3}{c}{\textbf{RF (physical target features \bm{$\mathcal{I}$})}}\\
\hline
$1$ & Number of trees & $1$, $50$, $100$, $250$, $500$, $1000$  \\
\multirow{2}{*}{$2$} & \multirow{2}{*}{\shortstack[l]{Nr. of features for best\\ split determination}} & \multirow{2}{*}{$1$, $3$, $5$, $7$, $9$, $11$}\\
&\\
\hline
\multicolumn{3}{c}{\textbf{LSTM (non-constant network target features \bm{$\mathcal{J}_{\text{nc}}$}})}\\
\hline
$1$ & \multirow{1}{*}{\shortstack[l]{Number of \gls{LSTM} layers}} & $1$, $2$, $3$\\
\multirow{1}{*}{$2$} & \multirow{1}{*}{\shortstack[l]{Batch size}} &$32$, $64$\\
\multirow{1}{*}{$3$} & \multirow{1}{*}{\shortstack[l]{Number of epochs}} &$100$, $200$, $500$\\
\multirow{1}{*}{$4$} & \multirow{1}{*}{\shortstack[l]{Dropout rate}} &$0$, $0.2$\\
\multirow{2}{*}{$5$} & \multirow{2}{*}{\shortstack[l]{Number of nodes in\\ \gls{LSTM} layers}} &\multirow{2}{*}{$20$, $50$, $100$}\\
&\\
\hline
\multicolumn{3}{l}{ $^{\mathrm{a}}$For other hyperparameters, default values from \cite{herzen2021darts} are used.} \\
\end{tabular}
\label{tab:RFR_hyperparameter}
\end{center}
\end{table}

\subsubsection{Automated model and detector tuning procedure}\label{sssec:modeling_procedure}
\paragraph{Methodology}
Implementing the signature extraction system requires performing the same automated tuning procedure for the detection and classification pipelines of all target features, which is schematically represented in Fig.~\ref{fig:Modeling_procedure}.
First step is the selection of forecasting models, which is concerned with the determination of appropriate target feature-specific hyperparameters.
The hyperparameters are selected based on a grid search using time series cross-validation on a training/validation set $X^c_{\text{train/val}}$ of normal operation data.
For each fold $X^c_{\text{fold}} = \{x^c_{1},x^c_{2},...,x^c_{N_{\text{fold}}} \ | \ x^c_{i} \in \mathbb{R} ~\forall i\}$ of the cross-validation, data is scaled to $\bar{x}^c_{i} \in [0,1]$ before training by
\begin{equation}
\bar{x}^c_{i} = \frac{x^c_i - \min \left({X^c_{\text{train}}}\right)}
{\max \left({X^c_{\text{train}}}\right) - \min \left({X^c_{\text{train}}}\right)}, \label{eq:scaling}
\end{equation}
$\forall x^c_{i} \in X^c_{\text{fold}}$, where $X^c_{\text{train}}$ corresponds to the first \SI{75}{\percent} of the respective fold.
In case a covariate time series $Z^c_{\text{train/val}}$ or multiple of them are used, they are scaled in the same way as target features in \eqref{eq:scaling}.
To finalize model selection, the resulting forecasting models are retrained on the respective full training/validation set $X^c_{\text{train/val}}$, again on values scaled according to \eqref{eq:scaling}, where now $X^c_{\text{train}}=X^c_{\text{train/val}}$.
\renewcommand{\thefigure}{7}
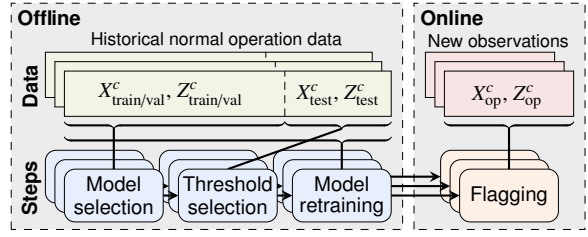
\begin{figure}[h]
  \centering
    \centering
\begin{tikzpicture} 

\draw[black, very thin, dashed, fill=gray!15] (-0.2-0.5,-0.1) rectangle (4.45,2.7+0.2);
\draw[black, very thin, dashed, fill=gray!15] (4.8-0.2,-0.1) rectangle (6.85,2.7+0.2);

\node[align=center,font=\sffamily] at (0.3-0.5,2.5+0.2) {\footnotesize \textbf{Offline}};
\node[align=center,font=\sffamily] at (5.3-0.2,2.5+0.2) {\footnotesize \textbf{Online}};

\node[align=center,font=\sffamily] at (2.1-0.1,2.2+0.2) {\scriptsize Historical normal operation data};
\node[align=center,font=\sffamily] at (5.81-0.1,2.2+0.2) {\scriptsize New observations};

\draw[black, very thin, fill=ggreen!15] (-0.05-0.2,1.45+0.2) rectangle (4.25-0.2,2.05+0.2);

\draw [line, black, densely dashed] (2.9-0.25,1.4+0.25) -- (2.9-0.25,2+0.25);

\draw[black, very thin, fill=ggreen!15] (-0.025-0.1,1.425+0.1) rectangle (4.275-0.1,2.025+0.1);

\draw [line, black, densely dashed] (2.9-0.125,1.4+0.125) -- (2.9-0.125,2+0.125);

\draw[black, very thin, fill=ggreen!15] (0,1.4) rectangle (4.3,2);

\draw[black, very thin, fill=rred!15] (4.95-0.2,1.45+0.2) rectangle (6.65-0.2,2.05+0.2);
\draw[black, very thin, fill=rred!15] (4.975-0.1,1.425+0.1) rectangle (6.675-0.1,2.025+0.1);
\draw[black, very thin, fill=rred!15] (5,1.4) rectangle (6.7,2);


\draw[black, very thin, fill=cornflowerblue!15, rounded corners] (-0.05-0.2,0.05+0.2) rectangle (1.25-0.2,0.75+0.2);
\draw[black, very thin, fill=cornflowerblue!15, rounded corners] (1.45-0.2,0.05+0.2) rectangle (2.75-0.2,0.75+0.2);
\draw[black, very thin, fill=cornflowerblue!15, rounded corners] (2.95-0.2,0.05+0.2) rectangle (4.25-0.2,0.75+0.2);
\draw[black, very thin, fill=sandybrown!15, rounded corners] (5.15-0.2,0.05+0.2) rectangle (6.45-0.2,0.75+0.2);
\draw [arrow, black, thick] (1.3-0.125,0.35+0.125) -- (1.5-0.125,0.35+0.125);
\draw [arrow, black, thick] (2.8-0.125,0.35+0.125) -- (3-0.125,0.35+0.125);
\draw [arrow, black, thick] (4.3-0.125,0.35+0.125) -- (5.2-0.125,0.35+0.125);
\draw [arrow, black, thick] (4.3-0.25,0.35+0.25) -- (5.2-0.25,0.35+0.25);

\draw[black, very thin, fill=cornflowerblue!15, rounded corners] (-0.025-0.1,0.025+0.1) rectangle (1.275-0.1,0.725+0.1);
\draw[black, very thin, fill=cornflowerblue!15, rounded corners] (1.475-0.1,0.025+0.1) rectangle (2.775-0.1,0.725+0.1);
\draw[black, very thin, fill=cornflowerblue!15, rounded corners] (2.975-0.1,0.025+0.1) rectangle (4.275-0.1,0.725+0.1);
\draw[black, very thin, fill=sandybrown!15, rounded corners] (5.175-0.1,0.025+0.1) rectangle (6.475-0.1,0.725+0.1);

\draw[black, very thin, fill=cornflowerblue!15, rounded corners] (0,0) rectangle (1.3,0.7);

\draw[black, very thin, fill=cornflowerblue!15, rounded corners] (3,0) rectangle (4.3,0.7);
\draw[black, very thin, fill=sandybrown!15, rounded corners] (5.2,0) rectangle (6.5,0.7);

\node[align=center,font=\sffamily] at (0.65,0.5) {\footnotesize Model};
\node[align=center,font=\sffamily] at (0.65,0.2) {\footnotesize selection};

\node[align=center,font=\sffamily] at (3.65,0.5) {\footnotesize Model};
\node[align=center,font=\sffamily] at (3.65,0.2) {\footnotesize retraining};

\node[align=center,font=\sffamily] at (5.85,0.35) {\footnotesize Flagging};




\node[align=center, rotate=90,font=\sffamily] at (-0.4-0.05,1.7+0.1) {\footnotesize \textbf{Data}};
\node[align=center, rotate=90,font=\sffamily] at (-0.38-0.05,0.35+0.1) {\footnotesize \textbf{Steps}};

\node[align=center] at (1.4,1.687) {\footnotesize $X^c_{\text{train/val}}$, $Z^c_{\text{train/val}}$};
\node[align=center] at (3.6,1.687) {\footnotesize $X^c_{\text{test}}$, $Z^c_{\text{test}}$};
\node[align=center] at (5.825,1.66) {\footnotesize $X^c_{\text{op}}$, $Z^c_{\text{op}}$};

\draw [arrow, black, thick] (1.3,0.35) -- (1.5,0.35);

\draw [arrow, black, thick] (2.8,0.35) -- (3,0.35);

\draw [arrow, black, thick] (4.3,0.35) -- (5.2,0.35);


\draw [line, black,thick] (0.65,0.7) -- (0.65,1.24);
\draw [line, black,thick] (2.15,0.7) -- (3.65,1.236);
\draw [line, black,thick] (3.65,0.7) -- (3.65,0.94+0.1);
\draw [line, black,thick] (5.85,0.7) -- (5.85,1.24);

\draw [line, black, densely dashed] (2.9,1.4) -- (2.9,2);

\draw [decorate, decoration = {calligraphic brace, raise=5pt, aspect=0.8489, mirror},thick] (0,1.3) --  (4.3,1.3);

\draw [decorate, decoration = {calligraphic brace, raise=5pt, aspect=0.2245, mirror},thick] (0,1.5) --  (2.9,1.5);

\draw [decorate, decoration = {calligraphic brace, raise=5pt, aspect=0.535, mirror},thick] (2.9,1.5) --  (4.3,1.5);

\draw [decorate, decoration = {calligraphic brace, raise=5pt, aspect=0.5, mirror},thick] (5,1.5) -- (6.7,1.5);



\draw[black, very thin, fill=cornflowerblue!15, rounded corners] (1.5,0) rectangle (2.8,0.7);

\node[align=center,font=\sffamily] at (2.15,0.5) {\footnotesize Threshold};
\node[align=center,font=\sffamily] at (2.15,0.2) {\footnotesize selection};

\end{tikzpicture}
  \caption{Automated tuning procedure of the anomaly detection and classification pipelines constituting the signature extraction system.} \label{fig:Modeling_procedure}
\end{figure}

After selecting appropriate forecasting models, the anomaly detectors (see Fig.~\ref{fig:Anomaly_detector}) of all pipelines are fitted to the respective target feature. 
For that purpose, a target feature-specific threshold $\tau_c$ is determined for each pipeline as follows.
First, the associated forecasting model is used to predict the expected values of a normal operation test set $X^c_{\text{test}}$ based on a rolling one-step ahead forecast.
The expected values are required to calculate all averaged distances of the test set $E^c_{\text{test}} = \{\varepsilon^c_{1},\varepsilon^c_{2},...,\varepsilon^c_{N_{\text{test}}} \ | \ \varepsilon^c_{i} \in \mathbb{R} ~\forall i\}$.
Based on $E^c_{\text{test}}$ and a threshold factor $f$, the feature-specific threshold is determined according to
\begin{equation}
    \tau_c = f \cdot \max\left(E^c_{\text{test}}\right). \label{eq:threshold fitting}
\end{equation}
If a target feature exhibits a high noise level or if non-optimal hyperparameters are selected, $\max \left(E^c_{\text{test}}\right)$ increases due to a weaker forecasting performance.
Thus, according to \eqref{eq:threshold fitting}, $\tau_c$ automatically adapts to the prediction performance of the respective forecasting model.
Objective of the proposed adaptive threshold is the reduction of \glspl{FP}.

Next, to fully exploit available historical data, the selected forecasting models are retrained on the respective totality of normal operation observations ($X^c_{\text{train/val}}+X^c_{\text{test}}$) and potentially covariates ($Z^c_{\text{train/val}}+Z^c_{\text{test}}$). 

After tuning the forecasting models and anomaly detectors, the detection and classification pipelines are ready for operation and can be applied to create anomaly flags for newly incoming observations $X^c_{\text{op}}$, which again are scaled according to \eqref{eq:scaling} with $X^c_{\text{train}}=X^c_{\text{train/val}}$.
Finally, the resulting anomaly flags of individual pipelines are grouped for each system zone of a \gls{CPS} to obtain event signatures as output of CyPhERS' Stage~1. 

\paragraph{Case-specific implementation} 
In the considered study case, $X^c_{\text{train/val}}$ and $Z^c_{\text{train/val}}$ are taken from the first \SI{75}{\percent} and $X^c_{\text{test}}$ and $Z^c_{\text{test}}$ from the remaining \SI{25}{\percent} of S0,  $\forall c \in \mathcal{I}$ and $\mathcal{J}$.
$X^c_{\text{op}}$ and $Z^c_{\text{op}}$ are provided by the three attack and fault scenarios S1-S3, $\forall c \in \mathcal{I}$ and $\mathcal{J}$.
Moreover, the threshold factor $f$ in \eqref{eq:threshold fitting} is specified to $f=1.5$.
Consequently, an anomaly within a target feature $c$ only is flagged if the average distance over the last 10 observations exceeds the biggest average distance during normal operation by at least \SI{50}{\percent}.
The target features are grouped according to the system zones depicted in Fig.~\ref{fig:Process_scheme}.

\setlength{\tabcolsep}{1.5pt}
\begin{table}[b!]
\fontsize{7.8}{9}\selectfont
\renewcommand*{\arraystretch}{1.2}
\caption{General anomaly flag interpretation rules for the definition of event signatures.}
\begin{center}
\begin{tabular}{l l}
\hline
\textbf{No.}&\multicolumn{1}{c}{\textbf{Rule description}}\\
\hline
\multirow{1}{*}{\shortstack[c]{1}} & \multirow{1}{*}{\shortstack[l]{Appearance of anomaly flags in a target feature indicates that the underlying physical or network component is affected (localization).}}\\
\multirow{1}{*}{\shortstack[c]{2}} & \multirow{1}{*}{\shortstack[l]{Anomaly flags exclusively in physical target features points towards a physical failure.}}\\
\multirow{1}{*}{\shortstack[c]{3}} & \multirow{1}{*}{\shortstack[l]{Anomaly flags exclusively in  network target features, including flags which indicate malicious activities\smash{$^{\mathrm{a}}$}, points towards a cyber attack.}}\\
\multirow{1}{*}{\shortstack[c]{4}} & \multirow{1}{*}{\shortstack[l]{Anomaly flags exclusively in network target features, without flags indicating malicious activities\smash{$^{\mathrm{a}}$}, points towards a network (device) failure.}}\\
\multirow{1}{*}{\shortstack[c]{5}} & \multirow{1}{*}{\shortstack[l]{Flags in both physical and network target features, including flags indicating malicious activities\smash{$^{\mathrm{a}}$}, points towards a cyber-physical attack.}}\\
\multirow{1}{*}{\shortstack[c]{6}} & \multirow{1}{*}{\shortstack[l]{Flags in ph. and netw. target features, w/o flags indicating mal. activities\smash{$^{\mathrm{a}}$}, indicate a network (device) failure entailing physical impact.}}\\
\multirow{1}{*}{\shortstack[c]{7}} & \multirow{1}{*}{\shortstack[l]{Anomaly flags exclusively in physical target features of one component indicates a local failure without impact on other components.}}\\
\multirow{1}{*}{\shortstack[c]{8}} & \multirow{1}{*}{\shortstack[l]{Physically plausible and coherent flags in physical target features of multiple components indicates a problem of their physical connection.}}\\
\multirow{1}{*}{\shortstack[c]{9}} & \multirow{1}{*}{\shortstack[l]{Anomaly flags exclusively in network target features of one device indicates a local device problem w/o impact on the rest of the system.}}\\
\multirow{1}{*}{\shortstack[c]{9.1}} & \multirow{1}{*}{\shortstack[l]{Rule 9 together with flags indicating malicious activity\smash{$^{\mathrm{a}}$} point towards a reconnaissance attack (e.g.,  scanning).}}\\
\multirow{1}{*}{\shortstack[c]{10}} & \multirow{1}{*}{\shortstack[l]{Flags simultaneously and exclusively in network target features of two netw. devices indicates a problem of their bilateral communication.}}\\
\multirow{1}{*}{\shortstack[c]{10.1}} & \multirow{1}{*}{\shortstack[l]{Rule 10 together with flags indicating malicious activities\smash{$^{\mathrm{a}}$} for both devices point towards a \gls{MITM} attack.}}\\
\multirow{1}{*}{\shortstack[c]{10.2}} & \multirow{1}{*}{\shortstack[l]{Rule 10.1 with flags in physical target features indicate a \gls{MITM} attack which manipulates process relevant data entailing physical impact.}}\\
\multirow{1}{*}{\shortstack[c]{11}} & \multirow{1}{*}{\shortstack[l]{Flags in physical target features indicating data disruption point towards disconnection of the network device which sends the data.}}\\
\multirow{1}{*}{\shortstack[c]{11.1}} & \multirow{1}{*}{\shortstack[l]{Rule 11 together with flags indicating malicious activities\smash{$^{\mathrm{a}}$} indicates a \gls{DoS} attack against the disconnected device.}}\\
\multirow{1}{*}{\shortstack[c]{11.2}} & \multirow{1}{*}{\shortstack[l]{Rule 11.1 with flags in ph. target features which indicate true physical impact point towards \gls{DoS} attack interrupting process relevant data.}}\\
\multirow{1}{*}{\shortstack[c]{12}} & \multirow{1}{*}{\shortstack[l]{Flags only in network target features of a device X and the ones connected to it indicate disconnection of device X.}}\\
\multirow{1}{*}{\shortstack[c]{12.1}} & \multirow{1}{*}{\shortstack[l]{Rule 12 together with flags indicating malicious activities\smash{$^{\mathrm{a}}$} point towards a \gls{DoS} attack ag. device X.}}\\
\multirow{1}{*}{\shortstack[c]{12.2}} & \multirow{1}{*}{\shortstack[l]{Rule 12.1 with flags in ph. target features which indicate true physical impact point towards \gls{DoS} attack interrupting process relevant data.}}\\
\hline
\multicolumn{2}{l}{$^{\mathrm{a}}$Malicious activities are, for example, communication with unknown devices or untypical connection requests from known devices.}\\
\end{tabular}
\label{tab:flag_interpretation_rules}
\end{center}
\end{table}
\renewcommand*{\arraystretch}{1}

\subsection{Signature evaluation (Stage~2)}\label{sssec:manual_reasoning}
\paragraph{Methodology}
Signature evaluation in CyPhERS is based on manually or automatically matching anomaly flags provided by Stage~1 with a database of known event signatures.
The possible attack and fault types that can affect a \gls{CPS} depend on the system's physical and digital components and architectures. 
Consequently, a set of event signatures which is valid across all systems cannot be defined.
However, some general anomaly flag interpretation rules which hold for most \glspl{CPS} have been identified, and are listed in Table~\ref{tab:flag_interpretation_rules}.
These general rules can be applied to create signature databases.
They range from simple principles, such as indication of affected system components through appearance of anomaly flags in the associated target features (rule 1), to more complex anomaly flag patterns which point to specific attack or fault types (e.g., rule 12.2).
Therefore, also signatures of different information detail can be defined, ranging from unclassified signatures, e.g. \texttt{Unknown event type affecting device X}, to specific event hypotheses, e.g. \texttt{\Gls{DoS} attack against device X entailing physical impact Y on component Z}.

\paragraph{Case-specific implementation}
Based on the general flag interpretation rules in Table~\ref{tab:flag_interpretation_rules}, a set of known event signatures is defined for the demonstration case.
Fig.~\ref{fig:Flag_patterns} visualizes them for selected victim devices, and on a reduced version of the system.
A description of the signatures is provided in Table~\ref{tab:flag_patterns_interpretation}.
For the sake of clarity, flags in network target features are grouped for each network device according to
\begin{equation}
    \Tilde{v}_t =   
\begin{cases}
    1 \text{ (\textit{anomaly})}& \text{if } \exists (v^c_t=2, 1, -1 \text{ or}-2) \in V^{\mathcal{J}}_t  \\
    0 \text{ (\textit{no anomaly})}& \text{otherwise}, 
\end{cases} \label{eq:anomaly_score_grouped}
\end{equation}
where $V^{\mathcal{J}}_t$ is the set of anomaly flags of one device at time $t$.
However, note that the grouped network target features of the victim devices during attacks must include flags indicating malicious activities (e.g., connection to unknown external device) to fulfill the attack signatures.
The differentiation between anomaly types is not applicable for the grouped anomaly flags $\Tilde{v}_t$ and thus neglected.
For the demonstration of CyPhERS on the given study case in Section~\ref{sec:results}, manual recognition of the defined event signatures is considered. 

\renewcommand{\thefigure}{8}
\begin{figure}[h]
\centering
    \input{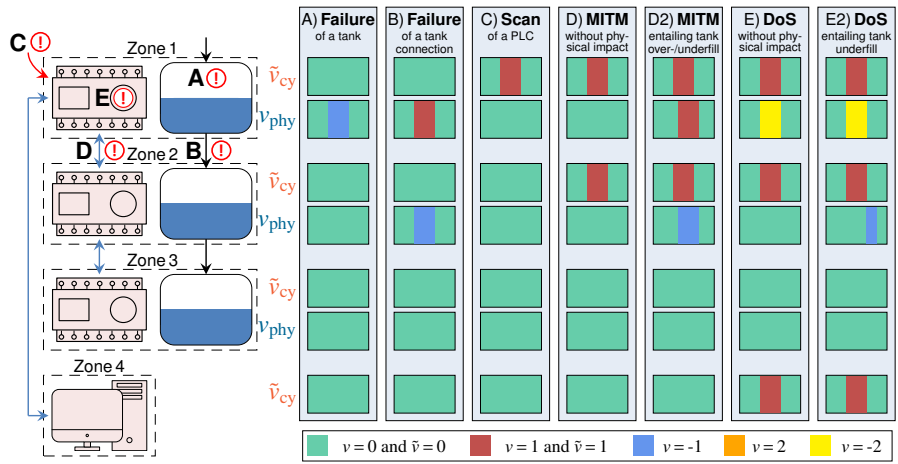}
    \caption{Event signatures for the study case. The signatures are depicted for selected victim devices and physical impacts. Grouped network features $\Tilde{v}_{\text{cy}}$ of the victim devices include flags indicating malicious activities during attacks.} \label{fig:Flag_patterns}
\end{figure}
\setlength{\tabcolsep}{1pt}
\begin{table}[h]
\fontsize{7.5}{9}\selectfont
\renewcommand*{\arraystretch}{1.2}
\caption{Description of the event signatures depicted in Fig.~8, and applied anomaly flag interpretation rules.}
\begin{center}
\begin{tabular}{l l l}
\hline
\textbf{Event signature}&\multicolumn{1}{l}{\textbf{Description}}&\textbf{Applied rules}\\
\hline
\multirow{1}{*}{\shortstack[l]{A)\,\textbf{Failure} of a tank}} & \multirow{1}{*}{\shortstack[l]{Abnormal high/low level of individual tank, no network anomaly. Thus, leak, in- or outflow failure\smash{$^{\mathrm{a}}$}}}. & \multirow{1}{*}{\shortstack[l]{1,2,7}}\\
\multirow{1}{*}{\shortstack[l]{B)\,\textbf{Failure} of tank connection}} & \multirow{1}{*}{\shortstack[l]{Parallel over-/underfill of linked tanks, no netw. anomaly.\,Thus, pump/valve failure or leak between them\smash{$^{\mathrm{a}}$}}}. & \multirow{1}{*}{\shortstack[l]{1,2,8}}\\
\multirow{1}{*}{\shortstack[l]{C)\,\textbf{Scan} of a \gls{PLC}}} & \multirow{1}{*}{\shortstack[l]{Individual \gls{PLC} affected, communication to unknown device with unusual TCP flags, no phy. anomalies.}}& \multirow{1}{*}{\shortstack[l]{1,3,9.1}}\\
\multirow{1}{*}{\shortstack[l]{D)\,\textbf{\Gls{MITM}} w/o phy. impact}} & \multirow{1}{*}{\shortstack[l]{Simult. \& excl. netw. anomalies in two conn. \glspl{PLC}, both comm. with unknown device, no phy. anomalies.}}& \multirow{1}{*}{\shortstack[l]{1,3,10.1}}\\
\multirow{1}{*}{\shortstack[l]{D2)\,\textbf{\Gls{MITM}} w. over-/underfill}} & \multirow{1}{*}{\shortstack[l]{Signature D with over-/underfill of tanks controlled by the victim \glspl{PLC}. Manip. fill levels distract pumps.}}& \multirow{1}{*}{\shortstack[l]{1,5,10.2}}\\
\multirow{2}{*}{\shortstack[l]{E)\,\textbf{\Gls{DoS}} w/o phy. impact}} & \multirow{2}{*}{\shortstack[l]{Network anomalies only for a device X (e.g., a \gls{PLC}) and the ones connected to it, physical data communi-\\cated by device X disrupted, connection of device X to unknown device, no phy. plausible anomalies.}}& \multirow{2}{*}{\shortstack[l]{1,3,11.1,12.1}}\\
&&\\
\multirow{1}{*}{\shortstack[l]{E2)\,\textbf{\Gls{DoS}} w. tank underfill}} & \multirow{1}{*}{\shortstack[l]{Signature E with underfill of a tank controlled by a \gls{PLC} which receives data from the disc. victim device.}}& \multirow{1}{*}{\shortstack[l]{1,5,11.2,12.2}}\\
\hline
\multicolumn{2}{l}{$^{\mathrm{a}}$Specific failure type can be concluded from anomaly flag directions and/or actuator type between tanks.}\\

\end{tabular}
\label{tab:flag_patterns_interpretation}
\end{center}
\end{table}
\renewcommand*{\arraystretch}{1}



\section{Demonstration} \label{sec:results}

In this section, results of applying CyPhERS on the three attack and fault scenarios (S1-S3) of the demonstration case are presented.
In preparation of that, Section~\ref{ssec:benchmarks} explains how the considered alternative approaches are represented for benchmarking, and Section~\ref{ssec:detailed_view_network_features} demonstrates attack signatures within ungrouped network target features.
Thereafter, S1-S3 are successively evaluated in the Sections \ref{ssec:evaluation_s1}-\ref{ssec:evaluation_s3}. 
In that context, examination of S2 includes comparison with the three benchmarks.
The investigated dataset contains some wrong ground truth event lengths and labels as well as further unlabeled anomalies.
Thus, focus of this section is on a qualitative demonstration of CyPhERS since a meaningful quantitative assessment is impractical under these circumstances. 

\subsection{Benchmark concepts} \label{ssec:benchmarks}
As part of the following demonstration of CyPhERS, a qualitative comparison to the existing event identification concepts introduced in Section~\ref{ssec:related_works} is conducted (group A-C). 
Group A is represented by considering only physical target features of CyPhERS, and grouping associated flags to provide system-wide monotypic anomaly flags $v_{\text{CPS}}$.
CyPhERS' physical target features without grouping their flags are considered for representing monotypic anomaly flags on feature level (group B). 
For group C, anomaly flags in both physical and network target features together are grouped, providing cyber-physical system-wide monotypic anomaly flags.

\subsection{Attack signatures in ungrouped network target features} \label{ssec:detailed_view_network_features}
The subsequent evaluation of S1-S3 considers grouped network target features for clarity (see Section~\ref{sssec:manual_reasoning}).
To demonstrate the appearance of the ungrouped flags that CyPhERS' Stage~1 provides during \gls{MITM}, \gls{DoS} and scanning attacks, they are depicted in 
Fig.~\ref{fig:Network_target_features_detailed} for selected network devices. 
\renewcommand{\thefigure}{9}
\begin{figure}[H]
  \centering
  \input{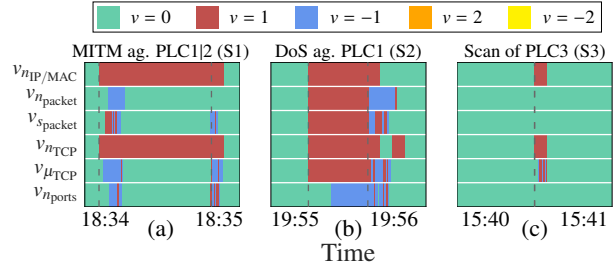}
    \caption{Flags in ungrouped network target features provided by CyPhERS' Stage~1 for a) PLC1 during \gls{MITM}, b) PLC1 during \gls{DoS} and c) PLC3 during scanning attack.} \label{fig:Network_target_features_detailed}
\end{figure}
During all three attack types malicious activities are flagged, which is a requirement of the pre-defined attack signatures (see Fig.~\ref{fig:Flag_patterns}).
These activities comprise communication with an unknown device ($v_{n_{\text{IP/MAC}}}=1$) and use of unusual TCP flags ($v_{n_{\text{TCP}}}=1$). 
Fig.~\ref{fig:Network_target_features_detailed} further indicates that the different attack types also express in distinctive signatures within the ungrouped network features.
For example, \gls{DoS} attacks result in pronounced anomaly flags on all features in contrast to the others.
Fig.~\ref{fig:MITM_detection} showcases this on a comparison to \gls{MITM} attacks. 
While the \gls{DoS} attack entails a global anomaly, the \gls{MITM} attack only results in a local one, primarily in the beginning and end of the attack\footnote{Note that the detection of the \gls{MITM} attack-induced local anomaly in Fig.~\ref{fig:MITM_detection} demonstrates the advantage of incorporating temporal information through time series models, as motivated in Section~\ref{sssec:TS_AD}.}.
While the differences within ungrouped network features are not taken into account in the following demonstration of CyPhERS, they potentially can be integrated to provide further information for distinction of different attack types. 
\renewcommand{\thefigure}{10}
\begin{figure}[h]
\centering
    \input{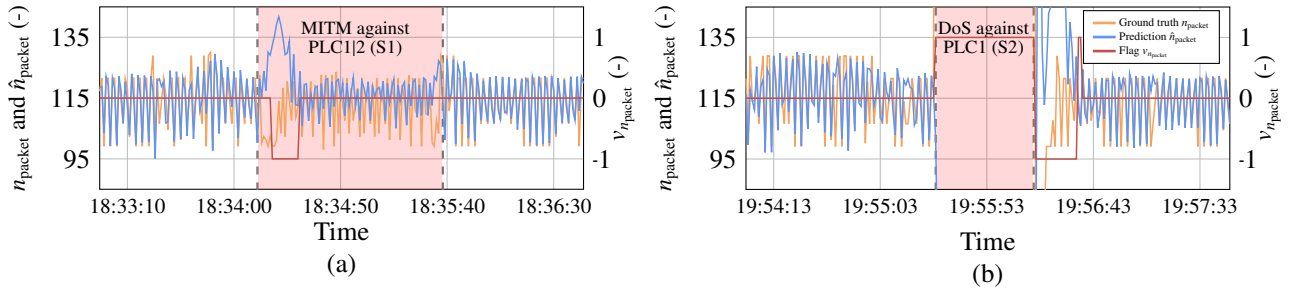}
    \caption{Anomalies in $n_{\text{packet}}$ of PLC1 induced by a (a) \gls{MITM} and (b) \gls{DoS} attack.} \label{fig:MITM_detection}
\end{figure}

\subsection{Evaluation of attack and fault scenario S1} \label{ssec:evaluation_s1}
The event signatures provided by CyPhERS' Stage~1 during S1 are depicted in Fig.~\ref{fig:S1}.
The anomaly flags of tank fill levels ($v_{H_{\text{T}1}}$-$v_{H_{\text{T}8}}$) are located next to the grouped network feature flags $\Tilde{v}$ of the \gls{PLC} controlling the respective process zone (see Fig.~\ref{fig:Process_scheme}).
The scenario comprises five \gls{MITM} attacks and three physical faults, affecting several network devices and physical components.
\renewcommand{\thefigure}{11}
\begin{figure}[t!]
  \centering
  \input{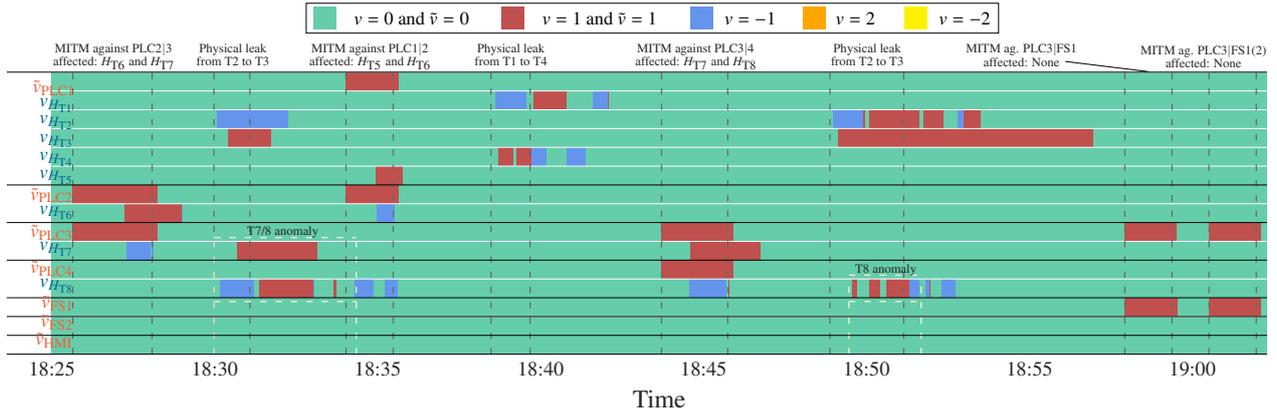}
  \caption{Event signatures of CyPhERS' Stage~1 during scenario S1. Additional anomalies not labeled by \cite{Faramondi2021} are marked using beige boxes, wrong ground truth labels are given in parenthesis.} \label{fig:S1}
\end{figure} 

\subsubsection{MITM attacks}
From Fig.~\ref{fig:S1} it is visible that anomaly flags during \gls{MITM} attacks either follow event signature D or D2 (see Fig.~\ref{fig:Flag_patterns}), which allows to conclude on the attack type, victim devices, attacker location and physical impact\footnote{According to \cite{Faramondi2021}, the last \gls{MITM} attack is supposed to affect \gls{PLC}3 and \gls{FS}2.
However, the anomaly flags point towards \gls{FS}1 instead of \gls{FS}2. A look on Fig.~\ref{fig:Process_scheme} shows that \gls{PLC}3 and \gls{FS}2 in fact do not communicate, proving successful identification of a wrong label.}.
For example, during the third \gls{MITM} attack ($\sim$18:45), signature D2 indicates a \gls{MITM} attack against \gls{PLC}3 and \gls{PLC}4 from an external device entailing overfill of T7 ($v_{H_{\text{T7}}}=1$) and underfill of T8 ($v_{H_{\text{T8}}}=-1$) due to failed pump activation as a result of manipulated fill levels exchanged between the victim \glspl{PLC}. 
Note that in all cases \gls{MITM}-induced anomalies are flagged in network features before the physical process is impacted, potentially allowing incident response mechanisms to take timely countermeasures.

The modification of the physical process during the first and third \gls{MITM} attack is comparatively strong.
As a result, the impact also extends over the next process cycle as well as consecutive tanks, which explains the two additional anomalies indicated in Fig.~\ref{fig:S1}. 
As an example, Fig.~\ref{fig:T7_add_anomaly} depicts the abnormal behavior of T7 in the next process cycle after the first \gls{MITM} attack.  
The comparison with the subsequent process cycle clearly indicates that T7 remains filled for an unusually long period. 
\renewcommand{\thefigure}{12}
\begin{figure}[h]
\centering
\input{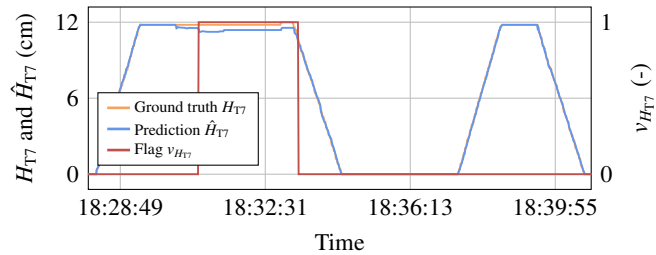}
\caption{Unlabeled anomaly of T7 during S1 as a result of the \gls{MITM} attack in the previous process cycle.} \label{fig:T7_add_anomaly}
\end{figure}

\subsubsection{Physical faults} \label{ssec:S1_physical_fault_detection}
Anomaly flags during physical faults follow event signature B (see Fig.~\ref{fig:Flag_patterns}), allowing to localize affected components, and infer fault types and physical impact.
For example, during the second fault ($\sim$18:40) the provided signature indicates a leak between T1 and T4, resulting in a simultaneous underfill of T1 ($v_{H_{\text{T1}}}=-1$) and overfill of T4 ($v_{H_{\text{T4}}}=1$). 
During the first and third fault, parallel anomalies from the previous \gls{MITM} attacks complicate recognition of signature B. 
However, as the flags for T7 and T8 can be explained by the preceding attacks, while a plausible connection to the abnormal behavior of T2 and T3 cannot be derived, these two events can be disentangled.

Fig.~\ref{fig:S1} indicates that CyPhERS' Stage~1 detects events at an early stage, and reliably differentiates anomaly types in the beginning of events.
However, the flags overrun the events for two reasons:
Firstly, the \gls{CPS} does not immediately recover after an attack or fault, thus, anomalies naturally persist.
Secondly, the anomaly detection and classification pipelines exhibit a \textit{recovering phase} after detected events.
As the detector evaluates the average distance of several consecutive observations, anomalies are flagged for some more steps even though the system behavior is already normal. 
Moreover, while passing an event, abnormal observations become the new model input, which manipulates the predictions entailing longer anomaly flags and unreliable flag types.
A concept improvement tackling this shortcoming is discussed in Section~\ref{ssec:concept_improvements}.

\subsection{Evaluation of attack and fault scenario S2} \label{ssec:evaluation_s2}
In the second scenario, \gls{DoS} attacks are performed in addition to \gls{MITM} attacks and physical faults.
The event signatures provided by CyPhERS' Stage~1 during S2 are depicted in Fig.~\ref{fig:S2} together with the ones of the three benchmarks.
\renewcommand{\thefigure}{13}
\begin{figure}[h]
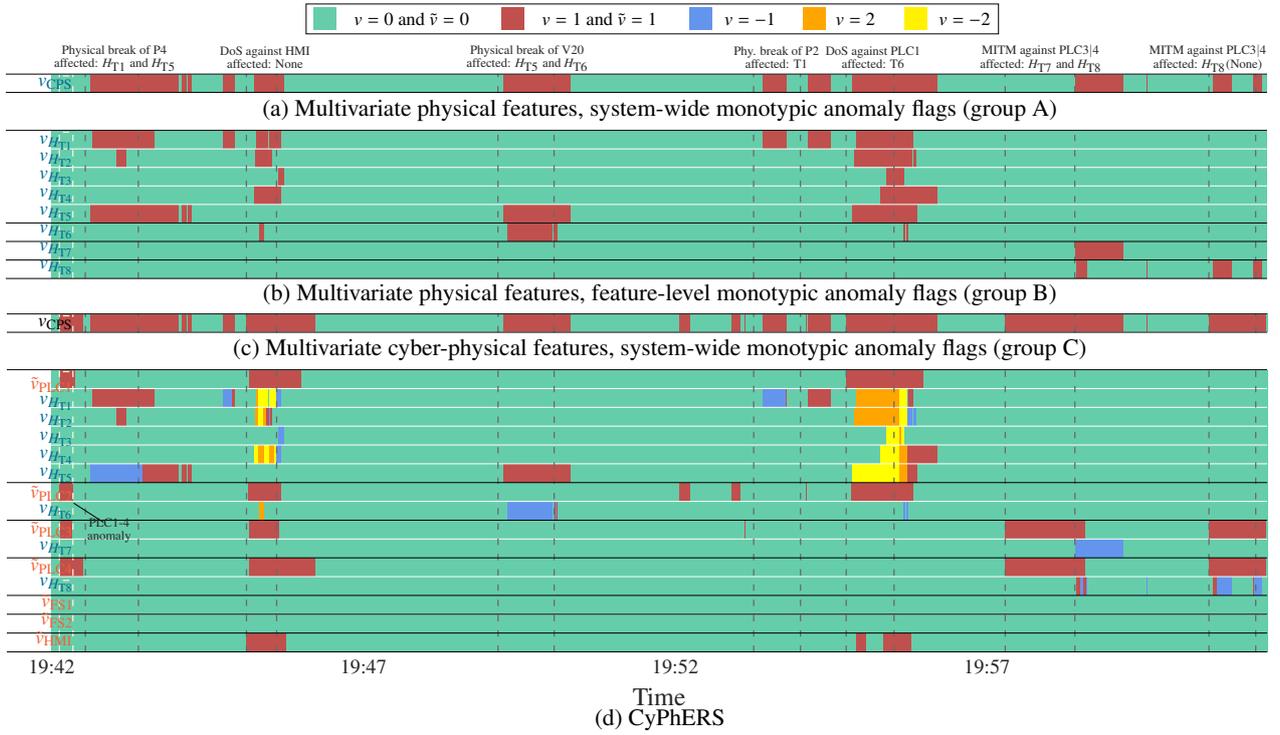

  \centering
  \hspace*{-0.09cm}\vspace*{-0.19cm}\input{Scenario_2_bm_systemwide.tex}
  \vspace*{-0.2cm} \hspace*{-0.09cm}\input{Scenario_2_bm_individual.tex}
  \vspace*{-0.17cm}\input{Scenario_2_bm_systemwide_cyphy.tex}
  \input{Scenario_2.tex}
    \caption{Event signatures of CyPhERS' Stage~1 and the three benchmarks during scenario S2. Additional anomalies not labeled by \cite{Faramondi2021} are marked using beige boxes, wrong ground truth labels are given in parenthesis.} \label{fig:S2}
\end{figure}

\subsubsection{MITM attacks and physical faults}
Anomaly flags provided by CyPhERS during faults either follow signature A or B. 
Thus, the affected tanks are localized, and pump or valve failures together with resulting physical impact inferred.
The benchmarks providing system-wide flags (Fig.~\ref{fig:S2} (a) and (c)) indicate fault-induced anomalies, however, do not provide information on affected components, event types and physical impact. 
In contrast, flagging anomalies in individual physical features (Fig.~\ref{fig:S2} (b)) additionally allows for localizing affected tanks. 
Nevertheless, the lack of network target features and anomaly type differentiation renders identification of event types and physical impact infeasible. 

Flags of CyPhERS during the two \gls{MITM} attacks follow signature D2, allowing to conclude attack type, victims, attacker location, and physical impact.
During the last attack, no physical impact should exist according to \cite{Faramondi2021}. 
However, as can be seen from Fig.~\ref{fig:T8_MITM_S2}, T8 in fact exhibits abnormal behavior. 
\renewcommand{\thefigure}{14}
\begin{figure}[h]
\centering
\input{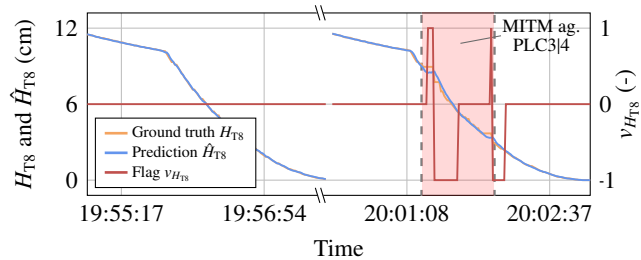}
\caption{Unlabeled anomaly of $H_{\text{T}8}$ in S2 during the \gls{MITM} attack against \gls{PLC}3 and \gls{PLC}4.} \label{fig:T8_MITM_S2}
\end{figure}

The benchmarks of group A and B only detect the physical impact of \gls{MITM} attacks (see Fig.~\ref{fig:S2} (a) and (b)). 
As a result, they exhibit high detection delays during the first one.
Moreover, none of the three benchmarks can indicate that anomalies are caused by a cyber attack.  

\subsubsection{DoS attacks}
The flags provided by CyPhERS' Stage~1 during the two \gls{DoS} attacks correspond to signature E or E2, allowing to conclude on attack type, victim device, attacker location, and physical impacts.
For example, in the second case, signature E2 indicates a \gls{DoS} attack against \gls{PLC}1 from an external device, resulting in a slight underfill of T6 ($v_{H_{\text{T6}}}=-1$) due to interrupted communication of $H_{\text{T}5}$ values from \gls{PLC}1 to \gls{PLC}2.

The three benchmarks neither indicate disconnection of a network device nor localize it, since they either only detect \gls{DoS}-induced anomalies in physical process data, or provide non-interpretable system-wide flags.
Moreover, as they only output monotypic flags, it cannot be inferred that most anomalies result from data disruption instead of true physical events.

\subsubsection{Unlabeled event}
CyPhERS' Stage~1 detects an unlabeled event in the beginning of S2, which is depicted in Fig.~\ref{fig:PLC2_flag_add_anomaly} on the example of \gls{PLC}2's $\mu_{\text{TCP}}$. 
Although the event signature is not known, fulfillment of flag interpretation rules 1 and 4 (see Table~\ref{tab:flag_interpretation_rules}) indicates a network failure only affecting \gls{PLC}1-4 without physical impact. 
This example demonstrates how CyPhERS provides real-time information including occurrence, affected devices, physical impact, and differentiation between network failure and cyber attack also for unknown event types.
The alternative detection strategies either entirely miss this event (see Fig.~\ref{fig:S2} (a) and (b)) or cannot provide any more information than its occurrence (see Fig.~\ref{fig:S2}~(c)).
\renewcommand{\thefigure}{15}
\begin{figure}[h]
\centering
    \begin{tikzpicture}
    \begin{axis}[
         width  = 0.48*\textwidth,
        height = 4cm,
        xlabel = {\small Time},
        ylabel={\small $\mu_{\text{TCP}}$ and $\Hat{\mu}_{\text{TCP}}$ (-)},
      legend image post style={scale=0.5},
      xtick pos=left,
       xticklabels={\small 19:42:30, \small 19:42:46, \small 19:43:02, \small 19:43:18},
       every tick label/.append style={font=\small},
        xmin=0,
        xmax=56,
        ymax=1.6,
        ymin=-0.6,
        xtick={4, 20, 36, 52},
        ytick={-0.5,0,0.5,1,1.5},
        yticklabels={0,0.15,0.3,0.45,0.6},
        ymajorgrids=true,
        xmajorgrids=true,
        ylabel shift = -0 pt,
        tick style={draw=none},
]

       

         \addplot[
    color=sandybrown,
    mark=none,
    line width=0.25mm,
    ]
    coordinates {(0, 0.98731714) (1, 0.46662822) (2, 0.57014084) (3, 0.99000007) (4, 0.23811767) (5, 0.78493077) (6, 0.98731714) (7, 0.0) (8, 0.99269754) (9, 0.99000007) (10, 0.02197015) (11, 0.9726431) (12, 0.8110193) (13, 0.32000002) (14, 0.89458567) (15, 0.61168903) (16, 0.49478996) (17, 0.8776152) (18, 0.5717514) (19, 0.57014084) (20, 0.89458567) (21, 0.34896553) (22, 0.7827624) (23, 0.89706373) (24, 0.11905883) (25, 0.99000007) (26, 2.1881082) (27, 5.782857) (28, 1.7771707) (29, 0.70490146) (30, 0.329711) (31, 0.9638096) (32, 0.6305618) (33, 0.40533334) (34, 0.99000007) (35, 0.40328768) (36, 0.6031111) (37, 0.99000007) (38, 0.34997118) (39, 0.6861017) (40, 0.99000007) (41, 0.118017495) (42, 0.89706373) (43, 0.99000007) (44, 0.11905883) (45, 0.86888885) (46, 0.75644445) (47, 0.34997118) (48, 0.89458567) (49, 0.5733711) (50, 0.57014084) (51, 0.89706373) (52, 0.4112321) (53, 0.7277995) (54, 0.98731714) (55, 0.23811767) (56, 0.7871111)};


    \addplot[
    color=cornflowerblue,
    mark=none,
    line width=0.25mm,
    ]
    coordinates {(0, 0.92210954) (1, 0.4634694) (2, 0.58136094) (3, 0.9553776) (4, 0.19652905) (5, 0.8359764) (6, 0.95711327) (7, 0.07982884) (8, 1.0568428) (9, 0.8480912) (10, 0.14972562) (11, 0.9432598) (12, 0.69821376) (13, 0.27800894) (14, 0.90927655) (15, 0.5160727) (16, 0.4787747) (17, 0.95448315) (18, 0.33086362) (19, 0.5531132) (20, 0.9510467) (21, 0.2361468) (22, 0.75274676) (23, 0.94085956) (24, 0.15555757) (25, 0.9607146) (26, 0.7937966) (27, -0.3808902) (28, -0.47752225) (29, 0.6965384) (30, 0.8876549) (31, 1.1245396) (32, 0.7001626) (33, 0.55703795) (34, 0.98061913) (35, 0.20174125) (36, 0.8567258) (37, 0.94359136) (38, 0.19804949) (39, 0.868861) (40, 0.9002611) (41, 0.16293329) (42, 1.0436301) (43, 0.8401012) (44, 0.22678941) (45, 0.9954386) (46, 0.6864707) (47, 0.37015057) (48, 0.9671495) (49, 0.46439976) (50, 0.5718617) (51, 0.975405) (52, 0.24272278) (53, 0.7215966) (54, 0.9495517) (55, 0.097393215) (56, 0.89146525)};
    
        
    \end{axis}

     \begin{axis}[
         width  = 0.48*\textwidth,
        height = 4cm,
        xlabel = {\small Time},
        axis line style={draw=none},
        hide x axis,
        ylabel={\small $v_{\mu_{\text{TCP}}}$ (-)},
      legend image post style={scale=0.5},
      xtick pos=left,
       xticklabels={\small 19:42:30, \small 19:42:46, \small 19:43:02, \small 19:43:18},
       every tick label/.append style={font=\small},
        xmin=0,
        xmax=56,
        ymax=2.2,
        ymin=-2.2,
        xtick={4, 20, 36, 52},
        axis y line*=right,
        ytick={-2,-1,0,1,2},
        yticklabels = {\scalebox{0.3}[1.0]{\( - \)}2,\scalebox{0.3}[1.0]{\( - \)}1, 0,1,2},
        ymajorgrids=false,
        xmajorgrids=false,
        ylabel shift = -5 pt,
        tick style={draw=none},
        legend pos=south west,
        legend cell align=left,
        legend style={nodes={scale=0.6, transform shape}, at={(0.011,0.025)},},
        legend image post style={line width =1pt}
]

            \addlegendentry{Ground truth \smash{$\mu_{\text{TCP}}$}}\addlegendimage{sandybrown}
        \addlegendentry{Prediction \smash{$\Hat{\mu}_{\text{TCP}}$}}\addlegendimage{cornflowerblue}
    
    \addplot[
    color=rred,
    mark=none,
    line width=0.25mm,
    ]
    coordinates {(0, 0) (1, 0) (2, 0) (3, 0) (4, 0) (5, 0) (6, 0) (7, 0) (8, 0) (9, 0) (10, 0) (11, 0) (12, 0) (13, 0) (14, 0) (15, 0) (16, 0) (17, 0) (18, 0) (19, 0) (20, 0) (21, 0) (22, 0) (23, 0) (24, 0) (25, 0) (26, 0) (27, 1) (28, 1) (29, 1) (30, -1) (31, -1) (32, -1) (33, -1) (34, 1) (35, 1) (36, -1) (37, 1) (38, 0) (39, 0) (40, 0) (41, 0) (42, 0) (43, 0) (44, 0) (45, 0) (46, 0) (47, 0) (48, 0) (49, 0) (50, 0) (51, 0) (52, 0) (53, 0) (54, 0) (55, 0) (56, 0)};

        \addlegendentry{Flag \smash{$v_{\mu_{\text{TCP}}}$}}
        
      
    
\end{axis}

\end{tikzpicture}
    \caption{Anomaly in $\mu_{\text{TCP}}$ of \gls{PLC}2 induced by an unlabeled event in S2.} \label{fig:PLC2_flag_add_anomaly}
\end{figure}
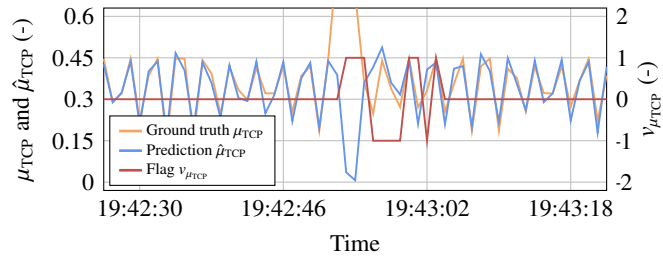

\subsection{Evaluation of attack and fault scenario S3} \label{ssec:evaluation_s3}
The third scenario adds scanning attacks to the previously evaluated attack and fault types.
The event signatures provided by CyPhERS’ Stage 1 during S3 are depicted in Fig.~\ref{fig:S3}.
\renewcommand{\thefigure}{16}
\begin{figure}[h]
  \centering
  \input{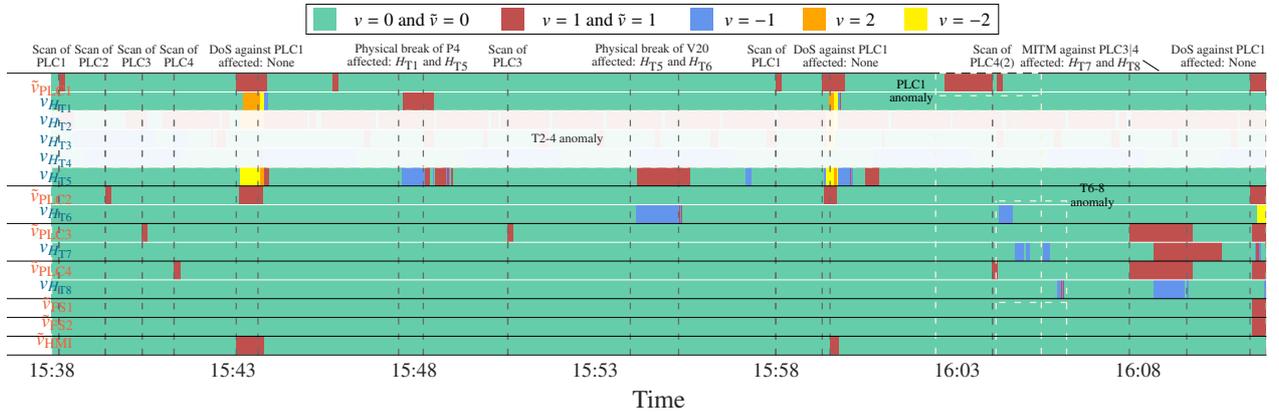}
    \caption{Event signatures of CyPhERS' Stage~1 during scenario S3. Additional anomalies not labeled by \cite{Faramondi2021} are marked using beige boxes, wrong ground truth labels are given in parenthesis.} \label{fig:S3}
\end{figure}
\renewcommand{\thefigure}{17}
\begin{figure}[t!]
\centering
    \input{add_anomaly_sc3_T2.tex}
    \caption{Unlabeled event of $H_{\text{T}2}$, $H_{\text{T}3}$ and $H_{\text{T}4}$ throughout S3 on the example of $H_{\text{T}2}$.} \label{fig:T2_sc3_add_anomaly}
\end{figure}

\subsubsection{Unlabeled events}
\paragraph{Event 1}S3 is characterized by a fundamental process modification not indicated by \cite{Faramondi2021}.
Throughout the entire scenario, anomalies are regularly flagged in $v_{H_{\text{T}2}}$, $v_{H_{\text{T}3}}$ and $v_{H_{\text{T}4}}$, as depicted in Fig.~\ref{fig:T2_sc3_add_anomaly} for $H_{\text{T}2}$.
While the event signature is not known, fulfillment of flag interpretation rules 1 and 2 (see Table~\ref{tab:flag_interpretation_rules}) points towards a physical failure only affecting the two sub-strings comprising T2-4 (see Fig.~\ref{fig:Process_scheme}). 
For the sake of clarity, $v_{H_{\text{T}2}}$, $v_{H_{\text{T}3}}$ and $v_{H_{\text{T}4}}$ are faded in Fig.~\ref{fig:S3}.

\paragraph{Event 2}CyPhERS' Stage~1 indicates two further unlabeled anomalies in S3, which likely result from the same event.
The anomaly in network traffic of \gls{PLC}1 is depicted in Fig.~\ref{fig:PLC1_payload_add_anomaly}.
Shortly after, a consecutive underfill of $H_{\text{T}6}$, $H_{\text{T}7}$ and $H_{\text{T}8}$ is indicated.
Compliance with flag interpretation rules 1 and 6 suggest a network device failure of \gls{PLC}1 potentially entailing the underfill of T6-T8 due to interrupted communication of $H_{\text{T}5}$ values from \gls{PLC}1 to \gls{PLC}2.


These two examples again showcase how CyPhERS provides real-time information for unknown event types including occurrence, affected devices, physical impact, and differentiation between physical failures, network failures, cyber attacks, and cyber-physical attacks.
\renewcommand{\thefigure}{18}
\begin{figure}[h!]
\centering
    \input{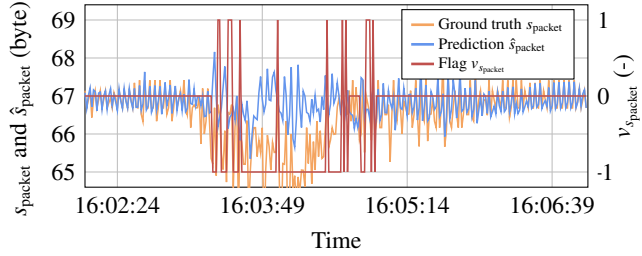}
    \caption{Unlabeled anomaly in $s_{\text{packet}}$ of \gls{PLC}1 in S3.} \label{fig:PLC1_payload_add_anomaly}
\end{figure}

\subsubsection{Scanning attacks}
Fig.~\ref{fig:S3} illustrates that anomaly flags during all scanning attacks follow signature C, which gives insights regarding attack type, victim, attacker location, and impact on the physical process\footnote{In this context, a wrong label is identified for the last scanning attack.}.
For example, during the first scan ($\sim$15:38), signature C, which includes flagging of communication with an unauthorized external device ($v_{n_{\text{IP/MAC}}}=1$) containing unusual TCP flags ($v_{n_{\text{TCP}}}=1$), indicates a scan of \gls{PLC}1 by an external device not impacting the physical process. 

\section{Discussion} \label{sec:discussion}
This section discusses key findings of the concept demonstration in Section~\ref{sec:results}.
In Section~\ref{ssec:proof_of_concept}, it is analyzed whether CyPhERS fulfils its intended purpose. 
Section~\ref{ssec:concept_improvements} addresses possible concept improvements.
Finally, Section~\ref{ssec:result_transferability} reflects on the transferability of the demonstration case results to other \glspl{CPS}.

\subsection{Proof of concept} \label{ssec:proof_of_concept}
The aim of CyPhERS is to provide \glspl{CPS} operators with relevant information on unknown and known types of attacks and faults for real-time incident response, while being independent of historical event observations.
The results in Section~\ref{sec:results} proof this thesis. 
All considered attack and fault types are identified, including localization of victim devices and attacker location as well as determination of the impact on the physical process, through matching anomaly flags provided by CyPhERS' Stage~1 with known event signatures.
Moreover, information on further unknown events, which are not officially labeled by the authors of the investigated dataset, are provided, including event occurrence, affected components, physical impact, and differentiation between physical failure, cyber attack, cyber-physical attack and network device failure. 

\subsection{Concept improvements} \label{ssec:concept_improvements}
One open issue identified by the concept demonstration is the recovering phase of the anomaly detection and classification pipelines.
Primary reason for this is the modification of the ground truth model inputs $x^{c}_{t-w},...,x^{c}_{t-1}$ in \eqref{eq:mapping_function} during and through anomalies, which affects the model's capability to predict the normal behavior of a target feature. 
To solve this issue $x^{c}_{t-w},...,x^{c}_{t-1}$ in \eqref{eq:mapping_function} could be replaced with the respective normal behavior predictions $\hat{x}^{c}_{t-w},...,\hat{x}^{c}_{t-1}$ during flagged anomalies.

Another improvement is seen in automating the process of creating the event signature database.
As of now, CyPhERS requires to define signatures by manually applying anomaly flag interpretation rules (see Table~\ref{tab:flag_interpretation_rules}) on the specific system at hand. 
In the future, this process should be automated through an application which generates event signatures by providing it with interpretation rules and specifications about the physical and digital components and architectures of a \gls{CPS}. 

\subsection{Result transferability to other \glspl{CPS}} \label{ssec:result_transferability}
As pointed out in Section~\ref{ssec:taxonomy}, process volatility and randomness may complicates application of CyPhERS to some \glspl{CPS}, including power system applications.
Since the study case considered in this work exhibits comparatively simple repeating process patterns, a feasibility demonstration of CyPhERS for systems with pronounced volatility and randomness is required. 
Moreover, to proof applicability for smaller \glspl{CPS} without dedicated human operator and for more complex processes where manual recognition of signatures would require a high cognitive effort, the automated signature evaluation in Stage~2 needs to be demonstrated.

\section{Conclusion and future work} \label{sec:conclusion}
This work introduces CyPhERS, a cyber-physical event reasoning system that provides real-time information about known and unknown types of attacks and faults in \glspl{CPS}, independent of historical event observations. 
CyPhERS uses a two-stage process to infer event information, including occurrence, location, root cause, and physical impact. 
In Stage~1, informative event signatures are created using methods such as cyber-physical data fusion, unsupervised multivariate time series anomaly detection, and anomaly type differentiation. 
In Stage~2, the event signatures are evaluated either automatically by matching with a signature database of known events or through manual interpretation by the operator. 
CyPhERS is demonstrated on a cyber-physical water distribution system, where it successfully identifies various attack and fault types, which includes localization of victim devices and attacker location as well as determination of attack or failure type and impact on the physical process. 
Additionally, CyPhERS provides information on unknown event types such as occurrence, affected components, physical impact, and differentiation between physical failure, cyber attack, and network failure. 
Future work will focus on demonstrating CyPhERS for systems with pronounced volatility and randomness under consideration of the automated signature evaluation in Stage~2.

\section*{Acknowledgement}
This work is partly funded by the Innovation Fund Denmark (IFD) under File No. 91363, and by the Helmholtz Association under the program ‘Energy System Design’.

\bibliographystyle{elsarticle-num} 
\bibliography{References}

\end{document}